\documentclass[11pt]{article}

\usepackage[T1]{fontenc}
\usepackage[utf8]{inputenc}
\usepackage{newtxtext,newtxmath}      % Times-like text + math (Science style)

\usepackage[letterpaper,margin=1in]{geometry}
\usepackage{setspace}
\usepackage{lineno}                    % continuous line numbers (enabled below)

\usepackage{graphicx}
\graphicspath{{figures/}}     % self-contained preprint: figures bundled alongside
\usepackage{amsmath}                   % (newtxmath supplies AMS symbols; no amssymb)
\usepackage{booktabs}
\usepackage{siunitx}
\usepackage{caption}                          % "Fig. 1 |" captions (Nature Portfolio style)
\DeclareCaptionLabelSeparator{natbar}{ $\vert$\ }
\usepackage{xcolor}
\usepackage{authblk}                          % author / affiliation block
\usepackage[super,sort&compress]{natbib}      % superscript citations (Nature Portfolio style)
\usepackage[colorlinks=true,allcolors=blue]{hyperref}

\newcommand{\beginsupplement}{%
  \setcounter{table}{0}\renewcommand{\thetable}{S\arabic{table}}%
  \setcounter{figure}{0}\renewcommand{\thefigure}{S\arabic{figure}}}

\title{\bfseries Optimal stimulation sites are not the most affected:
personalised models of resting-state fMRI in Alzheimer's disease}

\author[1,*]{Cristiano Capone}
\author[2]{Enza Cece}
\author[1]{Andrea Ciardiello}
\author[1]{Guido Gigante}
\author[2]{Evaristo Cisbani}
\author[3]{Maurizio Mattia}

\affil[1]{National Center for Radiation Protection and Computational Physics,
Istituto Superiore di Sanit\`a, 00161 Rome, Italy}
\affil[2]{Intelligenza Artificiale e Tecnologie Innovative per la Salute,
Istituto Superiore di Sanit\`a, 00161 Rome, Italy}
\affil[3]{Dipartimento di Neuroscienze, Istituto Superiore di Sanit\`a,
00161 Rome, Italy}
\affil[*]{Corresponding author: \href{mailto:cristiano.capone@iss.it}%
{cristiano.capone@iss.it}}
\date{}

\begin{document}
\maketitle

% ----------------------------------------------------------------------------
\begin{abstract}
\noindent
Resting-state functional connectivity (FC) is altered in Alzheimer's disease
(AD), widely regarded as a distributed network process; whether its signature
reduces to a few focal sites has not been tested causally, a question central to
targeted neuromodulation. We fit subject-specific, cross-subject-identifiable
models whose free-running dynamics reproduce those of each individual patient.
The fitted model parameters classify AD from controls at modest accuracy, below
that of structural atrophy; we build on the functional model nonetheless, because
dynamics, not tissue loss, are what stimulation can act on. Changing a virtual
patient's model connectivity toward the control template reverts its AD
classification, establishing in silico that the disease signature is correctable,
yet the required correction is \emph{intrinsically distributed}: a coordinated,
multi-site change of the model connectivity that no single-node edit reproduces.
Where, then, should a physically realisable focal drive act? A single-site drive
at the node whose connectivity is most altered fails to revert the classification
even at supra-physiological amplitudes, whereas selecting each patient's site by
its \emph{effect on the disease discriminant} achieves complete, individualised
reclassification from one site, and a real-time closed-loop controller reaches
comparable efficacy at lower dose using only causally available information.
Optimal targets are cortical and heterogeneous: the site to stimulate is not
where connectivity is most altered but where the network is most therapeutically
responsive.
\end{abstract}

% ============================================================================
\section*{Introduction}
Alzheimer's disease (AD) is increasingly understood as a disorder of brain
networks rather than of isolated regions, in which the progressive disruption of
large-scale functional interactions parallels cognitive decline~\cite{seeley2009,delbeuck2003}.
Resting-state functional connectivity (FC)
, the inter-regional correlation structure of spontaneous BOLD activity, is
among the most widely studied non-invasive readouts of this disruption, and AD
is consistently associated with altered connectivity, prominently within the
default-mode network~\cite{greicius2004,sheline2013}. FC has thus emerged as a
promising biomarker of Alzheimer's disease, although its diagnostic performance
and reproducibility remain insufficient for routine clinical
use~\cite{buckner2013}. Although AD is generally
viewed as a distributed, network-level disorder, a more basic question remains
open at the causal level: is the AD FC alteration \emph{reducible} to a small
number of regions or connections, or is it functionally irreducible to them,
distributed across the network?

The answer matters well beyond diagnosis. A focal alteration would, in
principle, offer a target for the emerging arsenal of network-level
interventions, deep-brain, transcranial-magnetic and transcranial
alternating-current stimulation, which are most effective when a small,
well-defined site can be engaged~\cite{fox2014}. A
distributed alteration, by contrast, would imply that no single target suffices.
This is fundamentally a causal, counterfactual question, what would happen to a
patient's connectivity if a given site were corrected, that group comparisons
of observational data cannot answer. Addressing it requires an individualised,
generative model of each subject's dynamics that can be perturbed in silico, in
the spirit of whole-brain network models and personalised brain
models~\cite{honey2009,deco2011,sanzleon2013,jirsa2017,wang2019}.

Reservoir computing offers a tractable route to such a model: a fixed random
recurrent network, driven by a subject's signals, provides a rich dynamical
basis from which a simple linear read-out can reconstruct that subject's
activity and connectivity~\cite{jaeger2001,maass2002}. Fitting the
read-out per subject yields a compact, individualised parameterisation of the
generative process, an individualised generative \emph{surrogate} of the
subject's FC, whose weights can be inspected, compared, and, crucially, modified.
This follows a broader programme in which models inferred from neural data
reproduce the data's own spontaneous spatiotemporal dynamics when simulated
freely~\cite{capone2018spont,capone2023simdata}, and in which network models
serve to study how large-scale dynamics shape function and
classification~\cite{capone2019slowosc}. Here we fit such models to
resting-state fMRI from AD patients and cognitively unimpaired controls, and
show that the closed-loop reconstructions are faithful to and identifiable from
individual connectivity, including its temporal (lagged) structure.

Building on this model, we develop the paper along a single arc, from reading
out the disease to intervening on it. First, we show that two complementary
read-outs of the fitted models, the geometry of the read-out weights across
patients and the lagged FC of the model's own reconstruction, classify AD from
controls with comparable, modest accuracy, robust to the choice of classifier
and dimensionality. We emphasise that functional connectivity is not the
strongest diagnostic read-out available: structural markers such as regional
grey-matter atrophy separate AD from controls more accurately and more
reproducibly across scanner sites (Supplementary Note~S1). Our purpose, however,
is not diagnosis but intervention, and atrophy is not a controllable variable,
one cannot stimulate lost tissue back. Stimulation acts on dynamics, so the
functional read-out, even as the weaker classifier, is the appropriate substrate
for a study of how activity can be causally reshaped; the FC-lag discriminant is
used here as a \emph{target} for perturbation, not as a diagnostic end in itself.
Second, we use the same models as an in-silico testbed and
address the stimulation question in two stages. Targeting the most-deviant
sites, the subcortical and limbic regions that diverge most from the control
template, fails to revert the classification at any amplitude, establishing that
the disease signature is not reducible to those focal nodes. Shifting the
selection criterion to the \emph{effect on the disease discriminant}, selecting,
for each patient, the site whose resonant drive most efficiently shifts the FC-lag
score toward the control distribution, achieves complete reclassification from a
single, personalised site, with closed-loop titration reducing the perturbation
cost further. The key finding is not merely that the signature is distributed,
but that the locus of maximal pathology and the locus of maximal therapeutic
effect dissociate: the former is subcortical and limbic; the latter is cortical
and patient-specific.

\section*{Results}

\subsection*{Dataset and functional-connectivity structure}
\begin{figure}[tbp]
  \centering
  \includegraphics[width=\linewidth]{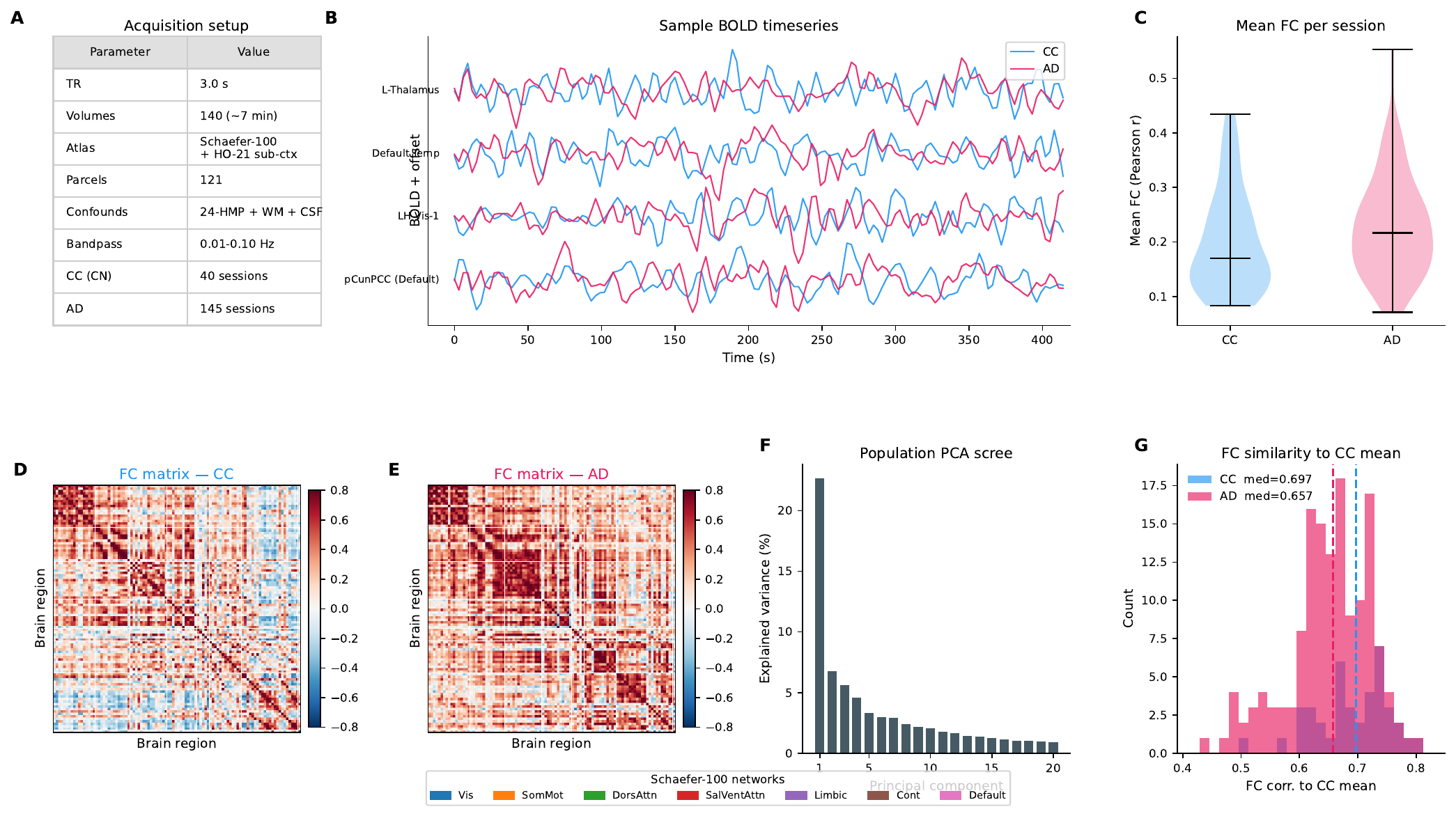}
  \caption{\textbf{Dataset and resting-state functional-connectivity structure.}
  (\textbf{A}) Acquisition and preprocessing parameters; cognitively unimpaired
  controls (CC) and Alzheimer's disease (AD) session counts.
  (\textbf{B}) Representative band-pass--filtered BOLD time series for four
  parcels in a CC (blue) and an AD (red) session.
  (\textbf{C}) Mean functional connectivity (FC; Pearson $r$ over parcel pairs)
  per session, CC vs AD. The two distributions do not differ significantly
  (two-sided Mann--Whitney $U$ over all $557$ CC and $145$ AD sessions,
  $p=0.08$; the panel shows a balanced $40$-session CC subsample), a weak trend
  toward higher mean FC in AD, consistent with the subtle, spatially diffuse
  group difference described below.
  (\textbf{D},\textbf{E}) Group-mean FC matrices (Schaefer-100 cortical parcels,
  network-sorted) for CC (D) and AD (E); white lines delimit the seven
  Yeo networks (legend).
  (\textbf{F}) Population-PCA scree: variance explained by the leading
  components of the pooled parcel covariance.
  (\textbf{G}) Per-session FC similarity (Pearson $r$ of the upper-triangular FC
  vector) to the CC-group mean FC, for CC and AD; dashed lines, group medians.}
  \label{fig:data}
\end{figure}

The cohort comprised resting-state fMRI sessions from cognitively unimpaired
controls (CC) and patients with Alzheimer's disease (AD), parcellated into
$121$ cortical and subcortical regions (Fig.~\ref{fig:data}A). At the level of
individual time series and their pairwise correlations the two groups were
broadly similar: representative BOLD traces (Fig.~\ref{fig:data}B), group-mean
FC matrices (Fig.~\ref{fig:data}D,E) and session-wise mean-FC distributions
(Fig.~\ref{fig:data}C) all shared the canonical resting-state network
organisation, and the population covariance was low-dimensional, a handful of
principal components capturing most of the variance (Fig.~\ref{fig:data}F).
Group differences were subtle and spatially diffuse rather than focal: AD
sessions were on average slightly less similar to the CC-group-mean FC than CC
sessions were, but the two distributions overlapped heavily
(Fig.~\ref{fig:data}G), so that no single connection or region trivially
separates the groups. This overlap motivated the model-based approach developed
below.

\subsection*{A subject-specific, identifiable reservoir model of FC}
\begin{figure}[tbp]
  \centering
  \includegraphics[width=\linewidth]{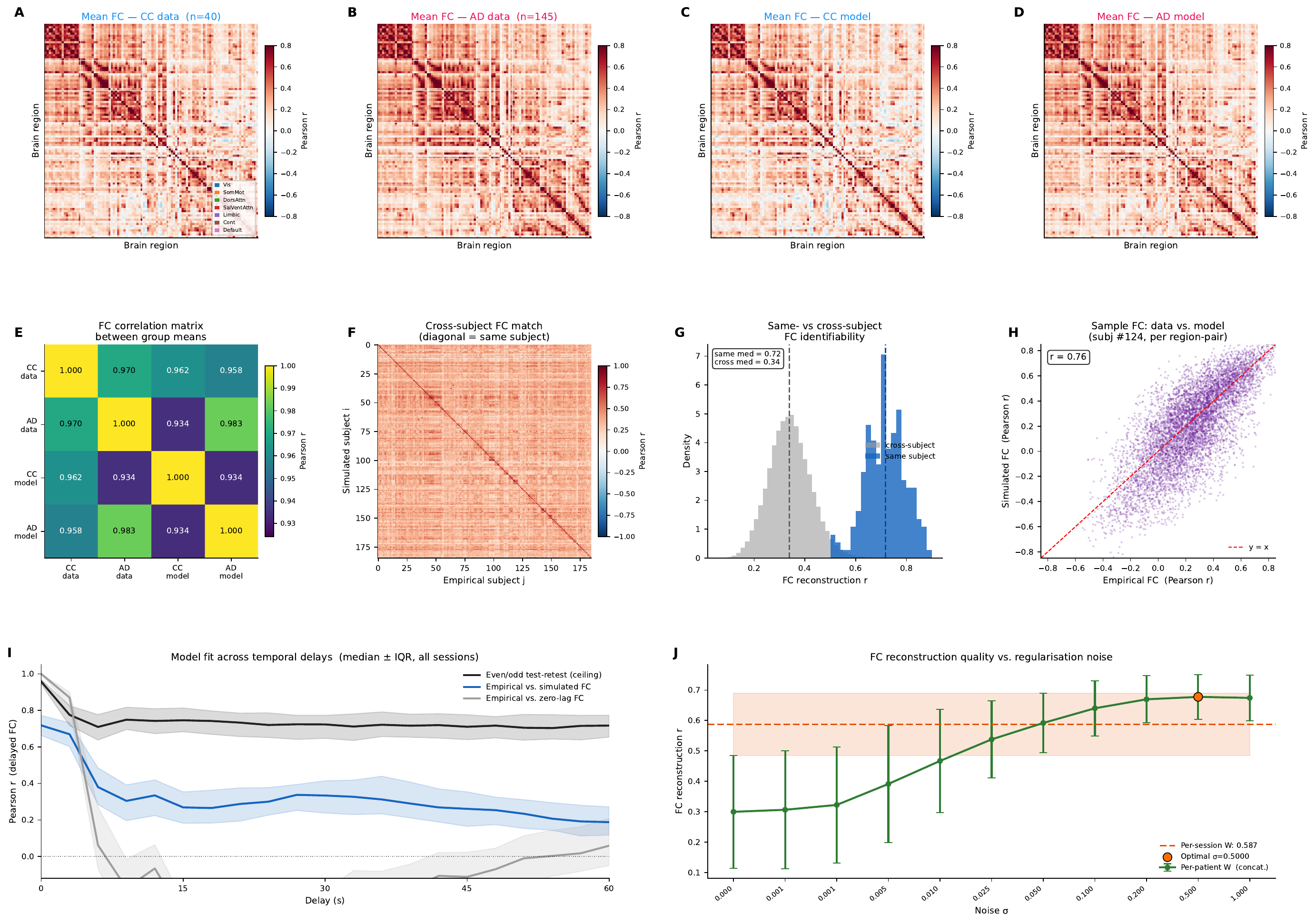}
  \caption{\textbf{A subject-specific reservoir model reconstructs and is
  identifiable from individual functional connectivity.}
  Per session, a fixed recurrent reservoir is teacher-forced on the
  PCA-projected BOLD; a linear read-out $W$ is fit, and the model is run
  closed-loop ($5$-step drive, then free-running) to generate a simulated
  reconstruction $Y=W^{\!\top}X$.
  (\textbf{A}--\textbf{D}) Group-mean FC for CC and AD, from data (A,B) and from
  the closed-loop model (C,D).
  (\textbf{E}) Correlation matrix (Pearson $r$) between the four group-mean FCs:
  each group's model FC most resembles its own data FC.
  (\textbf{F}) Cross-subject FC matrix: correlation between simulated subject $i$
  and empirical subject $j$; the diagonal (same subject) is the maximum of each
  row.
  (\textbf{G}) Distribution of same-subject (diagonal) vs cross-subject
  (off-diagonal) FC-reconstruction $r$; medians dashed.
  (\textbf{H}) Empirical vs simulated FC for a representative subject, one point
  per region pair ($r$ inset).
  (\textbf{I}) Model fit across temporal delays: correlation between empirical
  and simulated lagged FC (blue), empirical lagged vs zero-lag FC (grey), and
  even/odd test--retest ceiling (black); median $\pm$ IQR over sessions.
  (\textbf{J}) FC-reconstruction quality vs read-out regularisation noise
  $\sigma$ for per-patient $W$ (green); per-session baseline (orange); optimum
  marked.}
  \label{fig:model}
\end{figure}

For each session we fit a linear read-out from a fixed recurrent reservoir
driven by the individual's signals and ran the resulting model closed-loop to
generate a synthetic reconstruction (Methods). The model captured
individual connectivity well: group-mean reconstructed FC matrices
(Fig.~\ref{fig:model}C,D) reproduced the corresponding empirical matrices
(Fig.~\ref{fig:model}A,B), and across the four group means each model FC was
most similar to its own data FC, the AD model lying closest to the AD data and
the CC model to the CC data (Fig.~\ref{fig:model}E). The four group-mean matrices
are globally very similar (all pairwise $r>0.93$), so this ordering is only a weak
group-level effect; the decisive evidence for individual identifiability is the
within- versus across-subject contrast. Crucially, reconstructions
were subject-specific rather than generic: the correlation between a subject's
simulated and empirical FC was far higher within- than across-subjects
(Fig.~\ref{fig:model}F,G,H), establishing that the fitted models are
individually identifiable. The model also reproduced the temporal structure of
connectivity, the agreement between empirical and simulated lagged FC tracking
the empirical decay with delay and approaching the even/odd test--retest
ceiling (Fig.~\ref{fig:model}I), with reconstruction quality peaking at an
intermediate read-out regularisation and exceeding a per-session baseline
(Fig.~\ref{fig:model}J). Together these results establish the model as a
faithful, individualised generative account of each subject's
connectivity, the prerequisite for using it as an in-silico testbed.

\subsection*{Classification of AD from read-out geometry and reconstructed FC}
\begin{figure}[tbp]
  \centering
  \includegraphics[width=\linewidth]{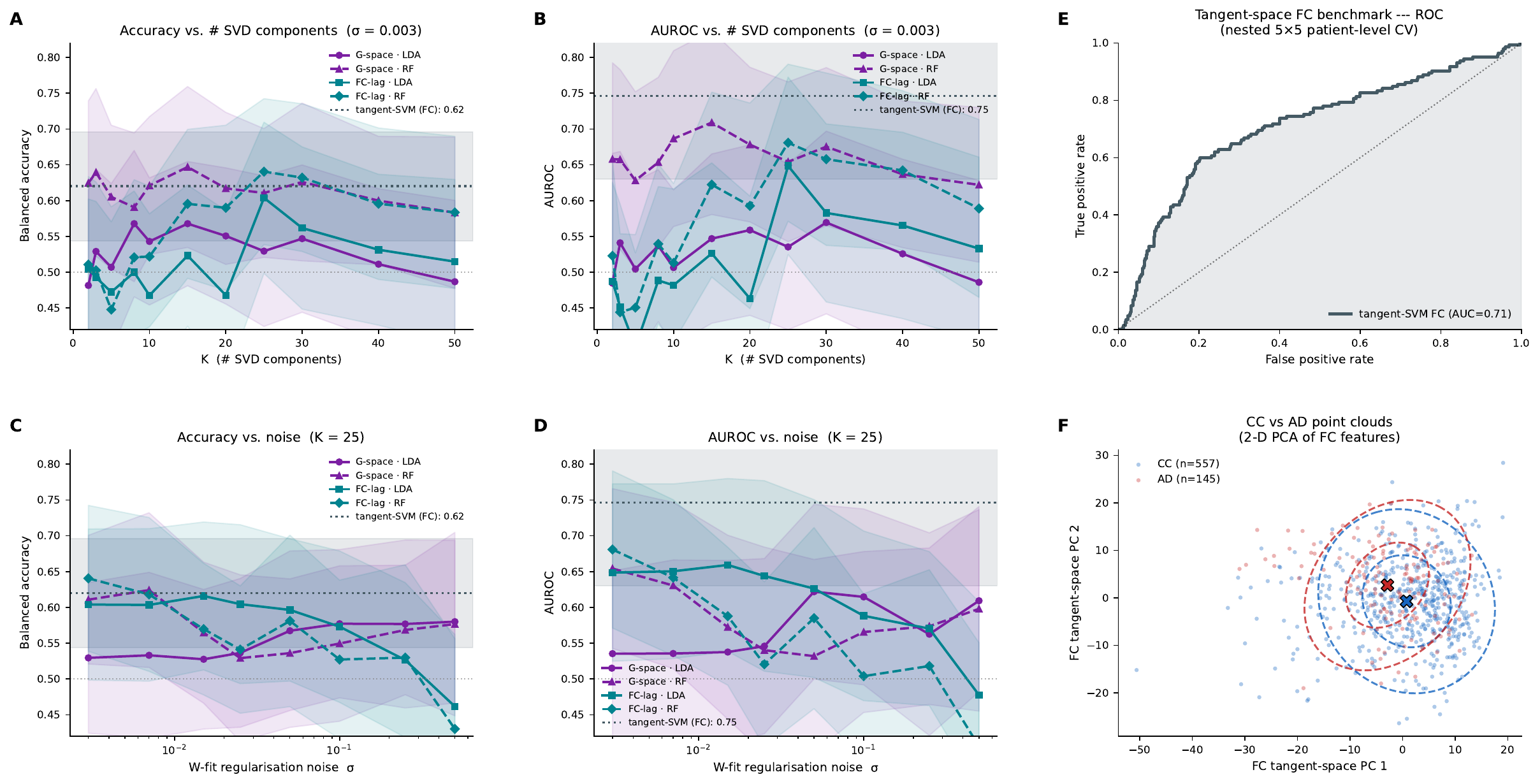}
  \caption{\textbf{CC-vs-AD classification: read-out geometry vs reconstructed
  functional connectivity.}
  Both classifiers derive from the \emph{same} per-patient read-out $W$:
  ``G-space'' uses the SVD geometry of the projected $W$ across patients;
  ``FC-lag'' uses the lagged-FC features (lags $0$--$2$) of the reconstruction
  $Y=W^{\!\top}X$. Each is classified with a balanced Fisher LDA and a random
  forest (RF), evaluated by repeated ($\times 10$) stratified $5$-fold
  ($80/20$) cross-validation. Lines, mean over folds; bands, $\pm 1$\,SD;
  colour, feature space; line style/marker, classifier.
  (\textbf{A},\textbf{B}) Balanced accuracy (A) and AUROC (B) vs the number of
  SVD components $K$ (at the reference noise $\sigma$).
  (\textbf{C},\textbf{D}) Balanced accuracy (C) and AUROC (D) vs read-out
  regularisation noise $\sigma$ (at the reference $K$).
  Grey dotted line, chance ($0.5$); dark dotted line with band, an independent
  tangent-space (Riemannian) FC $+$ linear-SVM classifier (AUROC $0.75\pm0.12$,
  mean $\pm$ SD over a \emph{nested} $5\times5$ patient-level CV with a train-only
  tangent reference and fixed, saved fold indices for reproducibility), a strong,
  leakage-free FC baseline. Both reservoir-based feature spaces plateau near
  $K\!\approx\!25$ and reach $\sim$0.70 AUROC, in the same modest range as the
  benchmark (at its lower end, within its wide spread), with G-space tolerating
  higher $\sigma$ whereas FC-lag favours low $\sigma$.
  (\textbf{E}) Out-of-fold ROC curve of the nested-CV tangent-space FC benchmark
  (AUC $\approx0.71$).
  (\textbf{F}) Two-dimensional PCA embedding of the tangent-space FC features,
  coloured by group (CC blue, AD red), with $1$- and $2$-SD covariance ellipses
  and group centroids ($\times$); the distributions overlap substantially,
  visualising the modest but above-chance separability that every classifier
  reaches.}
  \label{fig:classification}
\end{figure}

We next asked whether the fitted models carry diagnostic information, comparing
two read-outs of the same per-patient weights: the geometry of the read-out
across patients (``G-space'') and the lagged FC of the model's reconstruction
(``FC-lag'') (Methods). Both separated AD from CC well above
chance, but with modest and comparable performance, reaching $\sim$0.66--0.70
AUROC (Fig.~\ref{fig:classification}), in the range reported for FC-based AD
classifiers on comparable
cohorts~\cite{supekar2008,chen2011,khazaee2015,challis2015}. As a benchmark, an
independent tangent-space (Riemannian) FC classifier with a linear support-vector
machine, evaluated under \emph{nested} $5\times5$ patient-level cross-validation
with a train-only tangent reference, so that no test information enters the
feature construction, reached AUROC $0.75\pm0.12$ (out-of-fold $0.71$;
Fig.~\ref{fig:classification}, dotted band and ROC in
Fig.~\ref{fig:classification}E). The reservoir-based read-outs therefore lie in the
same modest range as this leakage-free FC benchmark, at its lower end but well
within its (wide) cross-validation spread, consistent with their being compact
per-subject parameterisations rather than classifiers optimised for discrimination.
As a further control we classified the \emph{experimental} lagged FC directly,
i.e.\ the lagged correlations of the empirical parcel BOLD ($143$ CC, $40$ AD
patients; same Gram-SVD embedding and leave-one-patient-out LDA/RF as the
model-based FC-lag): this reached comparable accuracy (AUROC $\approx0.71$,
peaking at $K\!\approx\!20$--$25$ and at lag~$2$), showing that reading out the
\emph{model's} reconstructed lagged FC neither adds nor loses diagnostic
information relative to the empirical lagged connectivity, but repackages the
same signal in the fitted, perturbable read-out. In a two-dimensional embedding of the FC features the CC and
AD point clouds overlap substantially, yet with distinct group centroids
(Fig.~\ref{fig:classification}F), consistent with the modest separability that
every classifier attains.
Discriminability increased with the
number of embedding components and plateaued near $K\!\approx\!25$, with little
benefit beyond (Fig.~\ref{fig:classification}A,B), and was insensitive to the
choice of classifier, a linear discriminant and a random forest performing
similarly. The two feature spaces differed mainly in their sensitivity to the
read-out regularisation: FC-lag was most accurate at low noise and degraded as
the read-out was over-regularised, whereas G-space was comparatively
noise-tolerant, remaining near its best over a broad range of $\sigma$
(Fig.~\ref{fig:classification}C,D). That two geometrically distinct read-outs
reach a similar ceiling is itself consistent with a modest, distributed disease
signal rather than a sharply localised one. Importantly, accuracy had not
saturated at the present sample size: a learning-curve analysis over the number
of patients used for fitting showed AUROC still rising with cohort size, and an
empirical extrapolation projects $\sim$0.76 AUROC at $N=100$ and $\sim$0.80 at
$N=200$ patients per group (Fig.~\ref{fig:scaling}), indicating that the modest
accuracy reflects sample size rather than a ceiling of the representation.

\subsection*{In-silico stimulation: pathology site versus therapeutic target}
\begin{figure}[tbp]
  \centering
  \includegraphics[width=\linewidth]{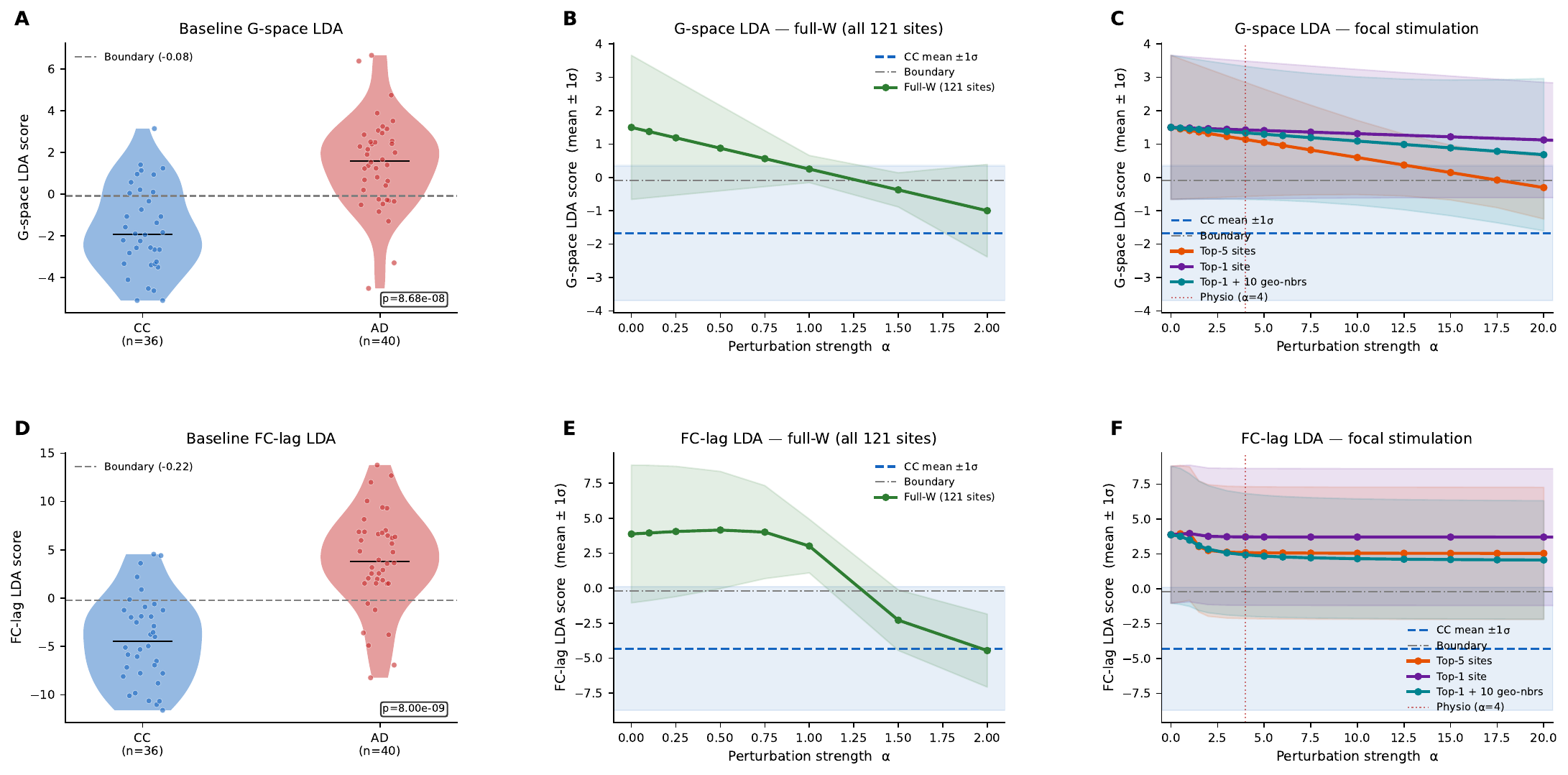}
  \caption{\textbf{In-silico stimulation: targeting the most-affected sites
  fails; only full-W correction reverts the classification.}
  AD read-outs are shifted toward the CC-mean read-out,
  $W_{\mathrm{int}}=(1-\alpha)\,W_{\mathrm{AD}}+\alpha\,\bar W_{\mathrm{CC}}$,
  either over all $121$ sites (full-W) or over focal site sets selected by
  $\lVert\Delta W\rVert$ (top-5, top-1, top-1$+$10 geometric neighbours).
  (\textbf{A}) Baseline G-space LDA scores for CC (blue) and AD (red) with the
  decision boundary (dashed); two-sided Mann--Whitney
  $p=8.7\times10^{-8}$ (also inset).
  (\textbf{B}) G-space LDA score vs $\alpha$ for full-W correction (mean $\pm1\sigma$
  over AD patients); CC mean $\pm1\sigma$ shaded; full-W enters the CC range at
  low $\alpha$.
  (\textbf{C}) Same for the three focal strategies (top-5, top-1,
  top-1$+$10 geometric neighbours); none crosses the boundary within
  physiological amplitude.
  (\textbf{D}--\textbf{F}) Analogous panels for the FC-lag LDA classifier
  (baseline two-sided Mann--Whitney $p=8.0\times10^{-9}$).
  The contrast between B/E (full-W succeeds) and C/F (focal fails at any
  amplitude) establishes that the AD FC signature is not reducible to the
  most-deviant sites.}
  \label{fig:stim}
\end{figure}

\begin{figure}[tbp]
  \centering
  \includegraphics[width=\linewidth]{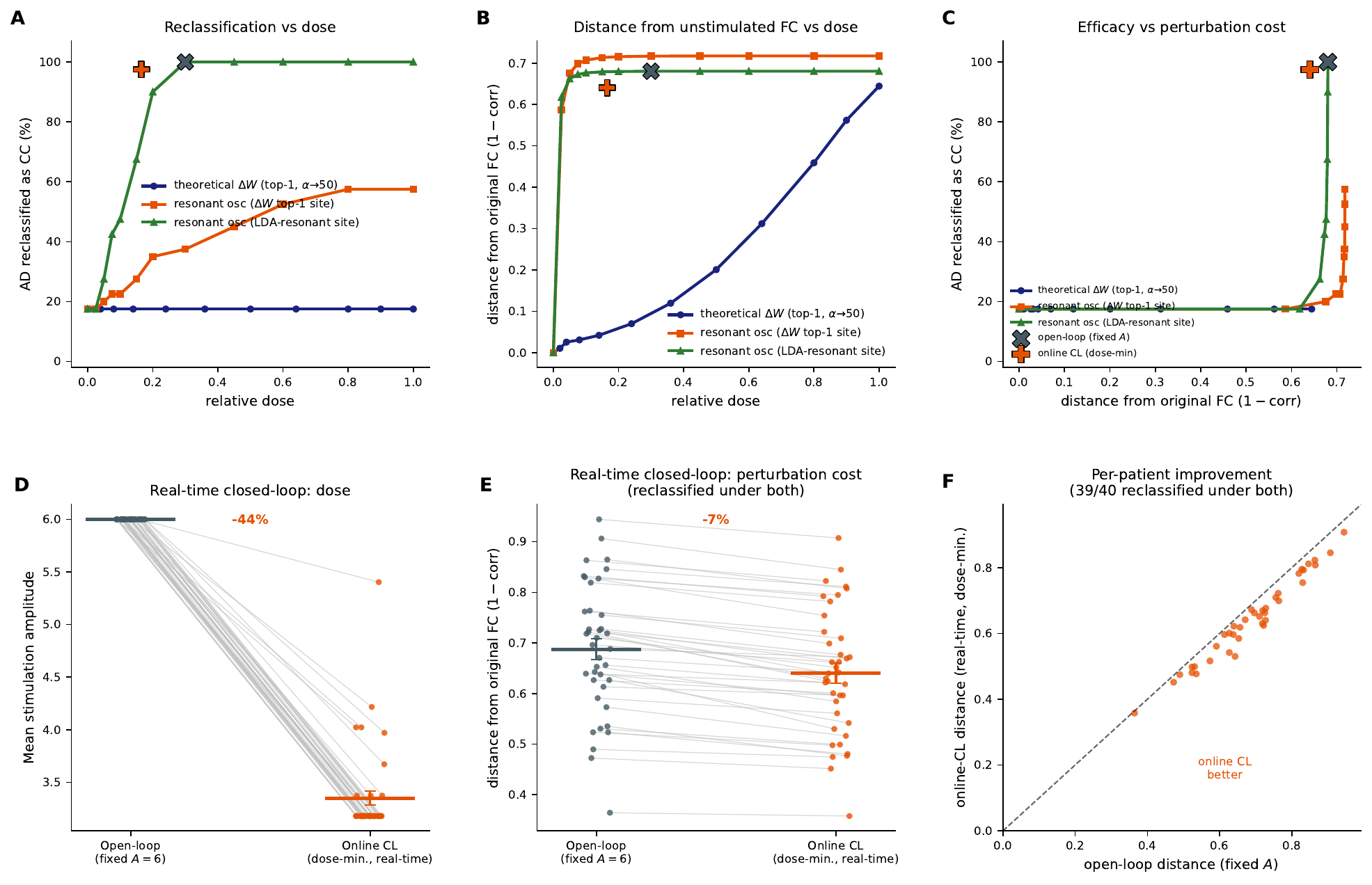}
  \caption{\textbf{Single-site stimulation and closed-loop control}
  ($N=40$ AD; distance $=1-$corr between the stimulated and unstimulated
  closed-loop free-run FC). \emph{Top row, single-site dose-response} comparing a
  \emph{theoretical} read-out correction ($\Delta W$ interpolation of the top-1
  most-affected column toward the CC mean, $\alpha\!\in\![0,50]$), a \emph{resonant}
  drive $A\sin(2\pi f_{\mathrm{eig}}t)$ at that same $\Delta W$ site, and a resonant
  drive at each patient's \emph{LDA-resonant} (discriminant-aligned) site
  ($A\!\in\![0,20]$).
  (\textbf{A}) Reclassification vs normalised dose: the single-column $\Delta W$
  correction is inert (flat at the $18\%$ baseline even at $\alpha=50$); the resonant
  drive at the $\Delta W$ site saturates near $\sim$$57\%$; the resonant drive at the
  LDA-resonant site reverts \emph{all} patients ($100\%$).
  (\textbf{B}) Distance from the original FC: the $\Delta W$ correction distorts only
  gradually, whereas both resonant drives jump to $\sim$$0.7$ at the smallest
  effective amplitude and saturate (the oscillation dominates the network-wide mode).
  (\textbf{C}) Efficacy vs perturbation cost: only the LDA-resonant drive converts
  perturbation into full reclassification.
  The closed-loop \emph{operating points} (patient means; grey $\times$ open-loop,
  orange $+$ online closed-loop) are overlaid on \textbf{A}--\textbf{C}: the online
  closed-loop point sits at a substantially smaller dose than the open-loop point,
  for a comparable, though not quite complete ($97.5\%$ vs $100\%$), reclassification
  rate and a modestly smaller distance.
  \emph{Bottom row, real-time closed-loop control at the LDA-resonant target} (the
  same curing site as the green curve of \textbf{A}), comparing a fixed
  \emph{open-loop} drive ($A=6$) to an \emph{online closed-loop} controller that
  titrates amplitude every few steps from a sliding-window estimate of the FC-lag
  biomarker computed only from the model's own past output, no oracle access to the
  future trajectory (Methods).
  (\textbf{D}) Per-patient stimulation amplitude (dose): online control lowers the
  mean dose from $6.0$ to $3.3$ ($\sim$$44\%$ lower) while nearly matching efficacy
  ($39/40$, $97.5\%$, vs $40/40$ open-loop).
  (\textbf{E}) Distance from baseline among patients reclassified under both
  protocols ($n=39$): online control lowers the mean distance by $\sim$$7\%$
  ($0.69\!\to\!0.64$).
  (\textbf{F}) Per-patient: most subjects lie below the diagonal, i.e.\ real-time
  control reaches nearly the same reversal at a smaller perturbation cost, the
  in-silico analogue of adaptive (closed-loop) neuromodulation, using only
  information available causally, online.}
  \label{fig:singlecompare}
\end{figure}

Finally, we used the model to ask whether the AD signature can be reverted by
intervention and, crucially, \emph{which} focal target, if any, makes a
single-site stimulation effective. We addressed this in two stages; as the
analysis below establishes, the sites where the AD read-out deviates most from
controls are not the sites where stimulation is most therapeutically effective.
Throughout, ``most-affected'' sites are those with the largest per-site read-out
deviation from the control template, i.e.\ the largest column norm
$\lVert\Delta\mathbf{W}\rVert$ of $\Delta\mathbf{W}=\bar{\mathbf{W}}_{\mathrm{CC}}
-\mathbf{W}_{\mathrm{AD}}$ (Methods). Shifting an AD
read-out toward the CC template over all regions (``full-W'') moved both
classifiers' scores into the CC range at small interpolation strength and
reclassified the large majority of AD patients as CC
for the G-space read-out and, identically, for the FC-lag read-out
(Fig.~\ref{fig:stim}A,B,D,E). In sharp contrast, applying the same correction to
focal site sets, the single most-affected site, the five most affected, or the
most-affected site together with its ten nearest neighbours, failed to revert
the classification within physiological amplitude limits, shifting only a
minority of patients even at much larger interpolation strengths, again in both
read-outs (Fig.~\ref{fig:stim}C,F). Physiological
diagnostics confirmed that the focal strategies were already operating at or
beyond plausible amplitudes: whereas the full-W correction left signal amplitude
near baseline, focal corrections required large local amplification to achieve
even partial effects (fig.~S10). An alternative focal
protocol, injecting an oscillation at the most-affected sites, tuned to the
reservoir's intrinsic frequencies, showed the same limitation, single- or
few-site stimulation moving the FC-lag score only weakly and only five-site
stimulation at supra-physiological amplitude approaching the decision boundary
(Fig.~\ref{fig:singlecompare}A; fig.~S7). Across both interpolation and oscillatory protocols,
no focal intervention reproduced the effect of the distributed correction,
indicating that, within the fitted models, the AD FC signature behaves as
functionally distributed and is not reduced to a small set of target sites.

\begin{figure}[tbp]
  \centering
  \includegraphics[width=\linewidth]{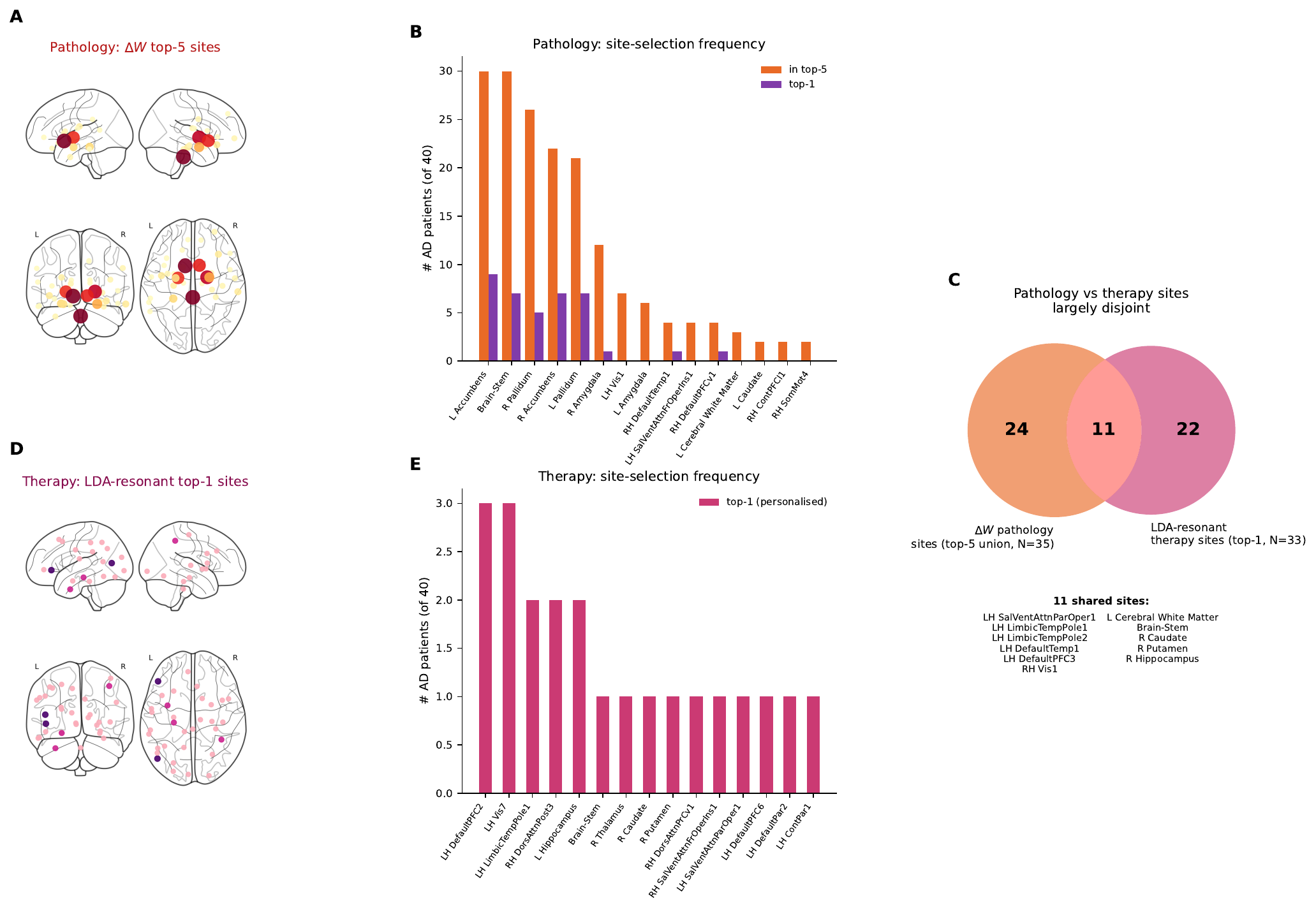}
  \caption{\textbf{Pathology sites and therapeutic-stimulation sites are
  largely disjoint.}
  \emph{Pathology} sites are ranked per AD patient by the per-site read-out
  correction magnitude $\lVert\Delta\mathbf{W}\rVert$ (column norm of
  $\Delta\mathbf{W}=\bar{\mathbf{W}}_{\mathrm{CC}}-\mathbf{W}_{\mathrm{AD}}$,
  top-5 per patient). \emph{Therapy} sites are each patient's personalised
  LDA-resonant target, the single site (top-1) whose resonant drive moves that
  patient's FC-lag discriminant most toward CC (Methods).
  (\textbf{A}) Glass-brain markers at the pathology sites (MNI centroids),
  coloured and sized by top-five selection frequency: deep subcortical/limbic
  location.
  (\textbf{B}) Pathology selection frequency: number of AD patients with each
  site among the top-five (orange) or as the single most-affected site (top-1,
  purple). The leading sites are subcortical and limbic, bilateral pallidum and
  nucleus accumbens, the brainstem and the amygdala.
  (\textbf{C}) Overlap between the two site sets: of the $35$ distinct
  pathology sites (top-5 union across patients) and $33$ distinct therapy sites
  (top-1 per patient), only $11$ sites ($\sim$$31\%$ and $\sim$$33\%$ of each
  set, respectively) are shared; the great majority of each set is exclusive to
  that criterion.
  (\textbf{D}) Glass-brain markers at the therapy sites, coloured and sized by
  top-1 selection frequency: predominantly cortical and, unlike the pathology
  sites, without a shared anatomical hub, each site is selected by at most three
  of the $40$ patients.
  (\textbf{E}) Therapy selection frequency: the leading therapy sites are the
  left Default-network prefrontal cortex and visual association cortex, with no
  site selected by more than three patients.}
  \label{fig:topsites}
\end{figure}

\begin{figure}[htbp]
  \centering
  \includegraphics[width=\linewidth]{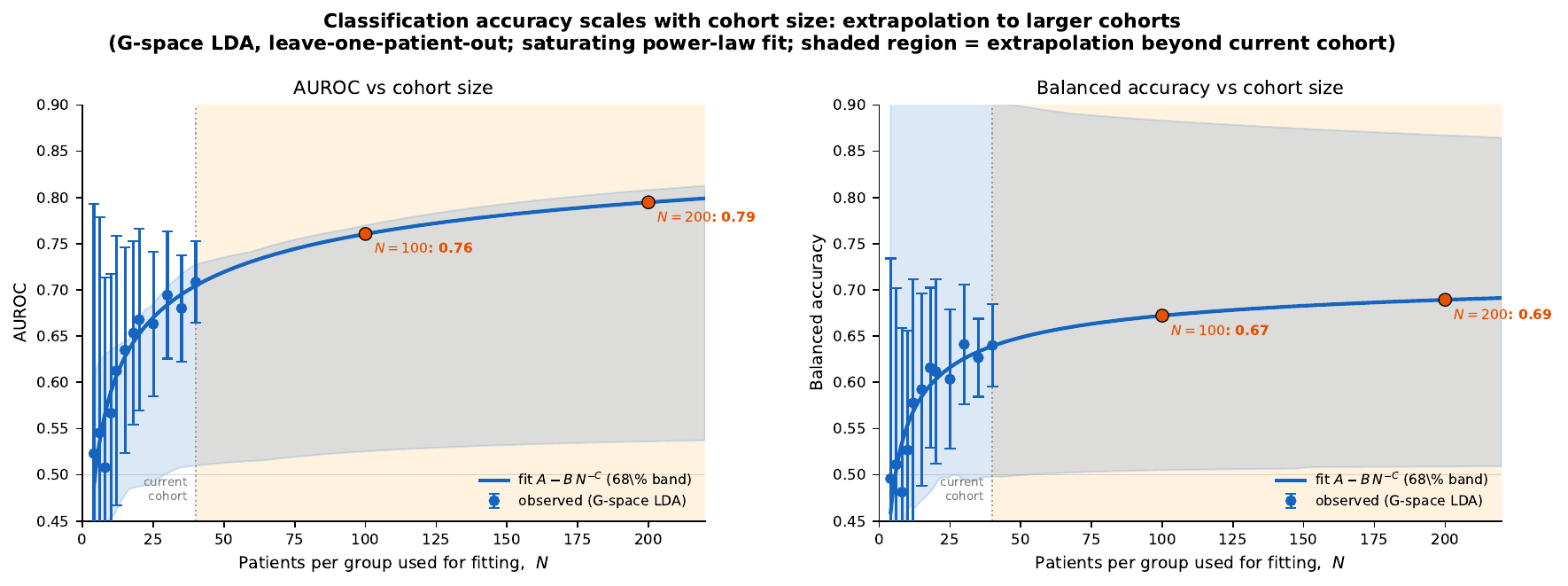}
  \caption{\textbf{Classification accuracy scales with cohort size.}
  G-space LDA balanced accuracy and AUROC (leave-one-patient-out) as a function
  of the number of patients per group used for fitting, $N$ (blue points, mean
  $\pm$ SD over resampled subsets). A saturating power law $m(N)=A-B\,N^{-C}$ is
  fit to each metric (blue line; shaded $68\%$ band from the fit covariance) and
  extrapolated beyond the current cohort (orange region). Accuracy has not
  saturated at $N=40$: the extrapolation projects AUROC $\approx0.76$ at $N=100$
  and $\approx0.80$ at $N=200$ (orange markers), indicating that the modest
  present accuracy is sample-size-limited rather than an intrinsic ceiling of the
  read-out representation.}
  \label{fig:scaling}
\end{figure}

The quantity $\lVert\Delta\mathbf{W}\rVert$ is best understood not as a localiser
of pathology but as an estimate of the change in the model's spatiotemporal
connectivity kernel required to map a patient's dynamics onto the control
template. As such it is intrinsically a \emph{distributed} object: it prescribes
a coordinated, multi-site, arbitrary spatiotemporal pattern of correction rather
than a single lesion coordinate. Our stimulation results make this concrete.
Implementing the full kernel change, interpolating all read-out columns toward
the control mean, reshapes the dynamics and reclassifies the large majority of
patients, whereas correcting a single column is inert
(Fig.~\ref{fig:singlecompare}): the correction cannot be reduced to one node. A
physically realisable focal oscillatory drive, by contrast, acts on a single
input site and is effective only where the network is dynamically responsive to
it, at sites selected by their effect on the discriminant, not by the magnitude
of their entry in $\Delta\mathbf{W}$. The dissociation between the two site sets
(Fig.~\ref{fig:topsites}) is therefore expected rather than paradoxical: a
distributed kernel correction and a focal resonant drive are different
interventions, and there is no reason their leading sites should coincide.

The anatomy of the largest kernel-correction weights is nonetheless interpretable,
with one caveat. The sites with the largest per-patient correction magnitude
$\lVert\Delta\mathbf{W}\rVert$ were consistently subcortical and limbic, most
frequently the bilateral pallidum and nucleus accumbens, the brainstem and the
amygdala (Fig.~\ref{fig:topsites}A,B). These are, however, also among the most
weakly inter-connected nodes of the parcellation, where a shared reservoir basis
constrains the per-subject read-out least; the ranking should thus be read as
indicating where the correction kernel carries its largest weights, not as an
independently validated map of Alzheimer pathology.

This failure of the pathology-guided approach motivated a change of selection
criterion. Rather than targeting the site of maximal read-out deviation, we
asked: which site, driven at the reservoir's resonant frequency, moves each
patient's FC-lag score most efficiently toward the control distribution? This
\emph{discriminant-aligned} target is a fundamentally different object from a
lesion marker, it is selected by its effect on the classifier, not by its own
magnitude of change, and identifying it requires the per-patient generative
model. The analysis below shows that this change of criterion is decisive: the
same type of focal, single-site, resonant intervention that fails at the
deviation-ranked target succeeds completely at the discriminant-aligned one.

Beyond identifying \emph{where} the disease signal concentrates, the model also
prescribes \emph{how} to drive a target. Efficacy is sharply frequency-tuned: a
sinusoidal drive at a focal site moves the FC-lag score toward the control
distribution most efficiently when its frequency matches the reservoir's dominant
eigenmode $f_{\mathrm{eig}}$, the resonant drive far outperforming an off-resonance
one per unit amplitude (fig.~S7A,B). This frequency, $f_{\mathrm{eig}}\approx0.077$
cycles/step, corresponds to $\approx0.026$\,Hz once expressed in physical time
through the TR ($=3$\,s), a unit conversion, not a dependence of the resonance on
TR, placing it in the well-established infraslow band in which resting-state
connectivity itself
fluctuates~\cite{fox2007,cordes2001,diantonio2026}, so the same fitted model
specifies both a location
(the subcortical/limbic targets of Fig.~\ref{fig:topsites}A,B) and an optimal driving
frequency, a spatial \emph{and} spectral prescription that descriptive analyses
cannot provide. This optimum is an \emph{interaction} of the two, however: a
resonant drive reshapes the FC network-wide, yet the effect still depends strongly
on the site, and the most disease-discriminative modes are not the most drivable
ones, so neither frequency nor site alone suffices (fig.~S6). As amplitude grows
the stimulated AD score shifts steadily toward the control range and a growing
fraction of patients cross the decision boundary (fig.~S7C,D). A full per-patient
amplitude$\times$frequency scan confirms that reclassification concentrates in a
narrow band at the resonance $f_{\mathrm{eig}}$, where patients cross the boundary
at the lowest amplitude and hence the smallest departure from the unstimulated
simulation, though reclassification and that departure trade off, so full reversal
still demands a large perturbation (fig.~S8). At
physiologically plausible amplitudes a focal drive does not, by itself, revert the
classification (Fig.~\ref{fig:stim}); its significance is methodological, demonstrating that an
identifiable generative model yields concrete, testable stimulation parameters. A
direct comparison of focal protocols on a common classifier shows that the
\emph{site-selection criterion} matters far more than the number of sites: a
2-site drive matched to the two leading reservoir eigenmodes and a single-site
resonant drive at the most-deviant site, selected for modal excitation and
disease norm respectively, both reclassify only $\sim$$50\%$ of patients.
Choosing the target instead by its \emph{effect on the discriminant}, the site
whose resonant drive moves the FC-lag score most toward CC, a criterion we call
\emph{LDA-resonant}, is markedly more effective: a single population-average such
site reclassifies $\sim$$65\%$ of patients, and personalising it per patient
reverts essentially all of them (Fig.~\ref{fig:singlecompare}A). These targets
are predominantly cortical, led by the left Default-network prefrontal cortex
(fig.~S3A,B), and highly heterogeneous, $33$ distinct sites for the $40$ patients
(fig.~S4), so no single global target is best for all. Tellingly, the
personalised LDA-resonant and the eigenmode-coupling drives perturb the
simulation \emph{near-identically} (correlation to the unstimulated model
$\approx0.32$--$0.34$; fig.~S5), yet only the discriminant-aligned target
converts that perturbation into full reclassification ($100\%$ vs
$\sim$$62\%$): selecting the site by its effect on the discriminant extracts more
therapeutic effect per unit of dynamical perturbation, and does so by perturbing
a single input site rather than rewriting all $121$ read-out weights. These
reversions still require supra-physiological amplitudes (fig.~S10); their
significance is methodological, showing that an identifiable per-subject model
turns ``where and how to stimulate'' into a directly computable prescription.

To isolate what makes a \emph{single-site} intervention work, we contrasted the
two ways of acting on one node (Fig.~\ref{fig:singlecompare}): the
\emph{theoretical} read-out correction, which interpolates that node's read-out
column toward the control mean, and the \emph{physical} resonant drive
($A\sin(2\pi f_{\mathrm{eig}}t)$). The single-column correction is essentially
inert: even at $\alpha=50$ it never reverts the classification (flat at the
$\sim$$18\%$ baseline) while progressively distorting the simulation, so a single
read-out weight has negligible leverage on the distributed FC-lag signature. The
resonant drive instead reclassifies by engaging the network-wide resonant mode,
but \emph{where} it is applied is decisive: at the $\Delta W$ most-affected
(subcortical) site it reaches only $\sim$$50$--$57\%$ even at high amplitude,
whereas at each patient's discriminant-aligned site it reverts \emph{all} of them
(Fig.~\ref{fig:singlecompare}A). Either way its distance from the original FC
jumps to $\sim$$0.7$ at the smallest effective amplitude and saturates
(Fig.~\ref{fig:singlecompare}B,C). At one site, then, a read-out edit is inert
and only a resonant drive at the right site cures, at a fixed, large perturbation
cost that the closed-loop control below is designed to minimise.

Having established that the therapeutic target is defined by its effect on the
discriminant rather than by pathological deviation, we can compare the two site
sets directly (Fig.~\ref{fig:topsites}C--E): the $35$ distinct sites ever
appearing in a patient's pathology top-5 and the $33$ distinct per-patient
therapy sites (personalised LDA-resonant targets) identified across the $40$ AD
patients share only $11$ sites, the great majority of each set is exclusive to
its own criterion, and the therapy sites, unlike the pathology sites, have no
shared anatomical hub.

\subsection*{Effectiveness of state-dependent perturbation at single-subject
level}

Because the fitted model is a closed-loop dynamical system, the same FC-lag score
that diagnoses a patient can serve as the feedback signal for a \emph{real-time,
causally valid} adaptive intervention: rather than an offline oracle with access to
the full future trajectory, we titrate the resonant drive's amplitude online, every
few steps, from a sliding window of the model's own past output (Methods). The
explicit window is a deliberately conservative choice: because the reservoir
satisfies the echo-state property, the relevant past is already embedded in its
instantaneous state, so an equivalent controller could read the biomarker
directly from $\mathbf{x}(t)$ without buffering past
output~\cite{albore2025search}.
At the fixed, personalised LDA-resonant site, this online controller
reclassifies $39/40$ ($97.5\%$) of AD patients, nearly matching the $100\%$ reached
by a fixed open-loop drive, while cutting the mean stimulation amplitude by
$\sim$$44\%$ ($A=6\to3.3$) and the departure from each subject's own dynamics by
$\sim$$7\%$ among patients reclassified under both protocols (distance
$0.69\!\to\!0.64$; Fig.~\ref{fig:singlecompare}D--F). Online amplitude titration
therefore converts the costly fixed drive into a substantially cheaper,
individualised protocol that reaches nearly the same in-silico reversal closer to
the patient's own resting dynamics, using only information available causally, in
real time.

A natural concern is that this reclassification could be inflated by fitting the
discriminant on the same patients it scores. We therefore re-evaluated the entire
pipeline under leave-one-AD-patient-out cross-validation: the FC-lag embedding,
the LDA discriminant, its decision threshold, \emph{and} the site-selection
criterion were all rebuilt from the training patients only and then applied to the
held-out patient (Methods; fig.~\ref{fig:lopo}). The therapeutic effect
survives. Honest cross-validation does raise the baseline (unstimulated)
misclassification of AD as CC from $17.5\%$ to $42.5\%$, so that only $23$ of the
$40$ AD patients lie on the AD side of the boundary without stimulation, consistent
with the modest classifier accuracy; we therefore quantify the effect as the net
fraction of these AD-side patients that stimulation moves across to the CC side. A
single \emph{population-average} resonant target, selected on the training patients
alone, achieves a leakage-free net cure of $\sim$$78\%$, and each patient's
\emph{personalised} LDA-resonant target moves essentially all AD-side patients
across the boundary (Methods). The single-site reclassification is therefore a
genuine, model-prescribed intervention effect rather than an artefact of
classifier leakage.

\section*{Discussion}
We fit identifiable, subject-specific reservoir-computing models of
resting-state functional connectivity and used them both to read out and to
intervene on the Alzheimer's signature. Two complementary read-outs of the
fitted models, the geometry of the read-out weights across patients and the
lagged FC of the model's own reconstruction, classified AD from controls with
comparable, modest accuracy. The central finding of the stimulation analysis is
a dissociation between the anatomy of pathology and the anatomy of effective
intervention. Targeting the sites of greatest read-out deviation from controls, 
predominantly subcortical and limbic, fails to revert the classification even at
supra-physiological amplitudes, confirming that the FC signature is not
concentrated in those regions. Selecting the target instead by its effect on the
disease discriminant under a resonant drive (the LDA-resonant criterion) achieves
complete reclassification from a single, individualised site, identifying for each
patient the node where stimulation is most therapeutically effective. The two site
sets are anatomically distinct: pathology concentrates in subcortical and limbic
hubs, whereas therapeutic leverage resides in cortical nodes, predominantly
prefrontal default-network and dorsal-attention cortex, that propagate network-wide
corrections through their connectivity. Whether this dissociation reflects the
biology of AD or the representational geometry of the model is a question our
framework raises but cannot, on its own, settle.

This conclusion is consistent with the view of AD as a network, or
``disconnection'', disorder rather than a regionally circumscribed
one~\cite{delbeuck2003,seeley2009}, and extends it with a causal, counterfactual
argument that observational group comparisons cannot make: within an
individualised generative model, targeting where the pathology is greatest is
not equivalent to targeting where stimulation is most effective. The practical
implication for neuromodulation is twofold. First, a single-target intervention
\emph{can} revert a distributed connectivity deficit, but only when the target
is selected for its effect on the disease discriminant rather than for its
pathological magnitude; this distinction is not accessible without a per-patient
generative model. Second, the effective site is highly patient-specific (33
distinct parcels across the 40 AD patients), so no single anatomical landmark
serves all individuals: personalised, model-informed targeting is required. Both
the per-patient discriminant-aligned sites and the population-average LDA-resonant
target (left prefrontal default-network cortex) are consistent with loci explored
in non-invasive AD neuromodulation~\cite{fox2014,koch2022}, lending translational
plausibility to the model-derived prescription. For patients whose optimal site
is inaccessible or whose response at a single site is insufficient, the present
framework also provides a way to design distributed, multi-site or network-shaped
stimulation strategies in silico before any intervention, complementing
model-based approaches that predict stimulation spread in whole-brain
models~\cite{spiegler2016,muldoon2016,kunze2016,deco2019}.

The anatomy of the read-out deviations is suggestive, though it should be read
with care. The regions whose read-out departed most from the control template
were predominantly subcortical and limbic, bilateral pallidum and nucleus
accumbens, brainstem and amygdala, structures of the limbic and basal-ganglia
circuitry; several of these overlap with regions harbouring early neurofibrillary
pathology in Braak staging~\cite{braak1991,braak2011}, although they are also the
most weakly inter-connected nodes of the parcellation, where the per-subject
read-out is least constrained, so this correspondence should be treated as
provisional rather than as a validated disease localiser. Regardless of that
interpretation, these maximally-deviant regions were not viable focal
intervention targets: the site of the largest single-node correction and the site
of greatest therapeutic leverage dissociate cleanly. A plausible mechanistic account is that pathology
originating in limbic hubs propagates along structural and functional connections,
distributing its imprint on connectivity across the network~\cite{raj2012,vogel2020};
the discriminative signal then resides in the network-wide pattern, so that
correcting the seed regions alone is insufficient. The cortical nodes that do
provide therapeutic leverage, prefrontal default-network, dorsal-attention and
visual cortex, are precisely those best positioned to redistribute a corrective
perturbation network-wide through their dense long-range connectivity, and they
interface with the subcortical structures of greatest pathological deviation via
established DMN--limbic pathways. This dissociation between lesion locus and
effective intervention site is consistent with the network-medicine principle
illustrated most clearly for deep-brain stimulation~\cite{fox2014}: the optimal
target is not the damaged node but the network node most effectively coupled to
the symptom circuit. The per-patient heterogeneity of the optimal sites (33
distinct parcels for 40 patients, Fig.~S4) further argues that no universal
anatomical landmark suffices, a result consistent with the clinical heterogeneity
of AD and with the emerging literature on connectivity-guided personalised
neuromodulation.

A distinctive feature of the framework is that it prescribes not only where but
how to stimulate. The driving frequency that most efficiently moved a target
toward the control distribution coincided with the model's intrinsic resonance
(fig.~S7), connecting our results to the broader principle of
frequency-specific neuromodulation. Stimulation efficacy is well known to depend
on frequency, and entraining endogenous rhythms is an active therapeutic
strategy in AD, most prominently $40$\,Hz ``gamma'' sensory and electrical
stimulation, which reduces pathology and improves function in mouse models and
early human studies~\cite{iaccarino2016,martorell2019}. Our prescription operates
at a far slower, infraslow timescale, the band in which resting-state
connectivity fluctuates, rather than in the gamma range, reflecting the
BOLD-derived nature of the model; we therefore make no claim about a specific
deliverable protocol. The conceptual contribution is that an identifiable,
subject-specific model turns ``at what frequency should one stimulate?'' into a
computable quantity tied to the network's own dynamics, complementing
connectome-model approaches that predict the spatial spread of
stimulation~\cite{spiegler2016,kunze2016}. A practical advantage is that, because
the reservoir dynamics and the read-out are \emph{linear}, the response to a
periodic drive, and hence the optimal stimulation site and frequency, follows
directly from the system's transfer function and eigenspectrum, rather than from
an exhaustive search: the resonance we observe is precisely the peak of this
linear frequency response. The model thus affords a closed-form, analytically
tractable prescription of optimal stimulation parameters. Notably, this optimum is
an \emph{interaction} of site and resonance that no single-factor heuristic
captures: a resonant drive reshapes the FC network-wide, yet the reclassification
still depends strongly on the site (fig.~S6A), the disease-discriminative modes are
not the drivable ones (the dominant, most-drivable mode is only weakly
discriminative; fig.~S6B), and steering by a discriminative-but-undrivable
frequency, alone or combined with the resonance, does not help (fig.~S6C,D).
Selecting the target by its \emph{effect on the discriminant under a resonant
drive} (the LDA-resonant criterion) is what jointly resolves drivability and
direction, and is why it outperforms modal-coupling or disease-deviation selection.
Reassuringly, when the
target is selected for its effect on the disease discriminant, the leading
``drive-toward-health'' site falls in the Default-network prefrontal cortex
(fig.~S3), a hub of the network most consistently disrupted in
AD~\cite{greicius2004,sheline2013} and a prime locus for non-invasive
neuromodulation, e.g.\ precuneus/Default-network stimulation
trials~\cite{koch2022}, lending biological plausibility to the model-derived
prescription. The heterogeneity of the per-patient optimal targets further argues
that effective targeting must be individualised rather than one-size-fits-all,
consistent with connectivity-guided personalised stimulation~\cite{fox2014}.
Because the fitted model is a dynamical system, the same biomarker that scores a
patient can also \emph{close the loop} in real time: titrating the drive's
amplitude online against a sliding-window FC-lag estimate, using only the model's
own past output rather than an offline oracle, reclassifies $97.5\%$ of patients
(nearly matching the $100\%$ open-loop rate) while cutting both the stimulation
dose ($\sim$$44\%$) and the departure from their own dynamics ($\sim$$7\%$ among
patients reclassified under both protocols) relative to a fixed open-loop drive
(Fig.~\ref{fig:singlecompare}), the in-silico counterpart of adaptive
(closed-loop) neuromodulation, in which a causally available, minimal-dose
perturbation that crosses the diagnostic boundary, rather than a
supra-physiological open-loop drive, is sought. We emphasise, however, that these are \emph{model-level} prescriptions:
the reservoir is fit to infraslow BOLD connectivity, not electrophysiology, so an
``intervention'' here is an abstract input to the regional FC dynamics, and its
optimal spatial pattern and (infraslow) frequency should be read as targets at the
level of regional drive, not as cellular or oscillatory mechanisms. With that
caveat, the focal, slowly-modulated character maps in spirit onto closed-loop,
biomarker-guided neuromodulation; translating it to any specific
technology, and validating it against electrophysiological or longitudinal
outcomes, is left to future work.

More broadly, the study illustrates a general methodology: an identifiable
individualised generative surrogate of a subject's dynamics, fit once, can serve
simultaneously as a biomarker substrate and as a counterfactual testbed. Because
the read-out is a compact per-subject parameterisation, the same fitted object
that classifies the disease can be perturbed to ask what would reverse it, a
loop from decoding to intervention that is difficult to close with descriptive
analyses alone, and that transfers directly to other conditions and to
stimulation planning.

Several limitations qualify these conclusions. The cohort is cross-sectional and
of modest size, and the classifiers, while honestly cross-validated, reach only
moderate accuracy and embed each subject relative to the group (a transductive
embedding); per-subject read-outs were estimated from limited data, adding
fitting noise that we characterised but did not eliminate. The therapy-target
reclassification is not, however, an artefact of this transductive fit: it
survives a fully leave-one-AD-patient-out reconstruction of the embedding,
discriminant, threshold and target selection, once the effect is read as the net
fraction of AD-side patients moved to control, above an honest (higher) baseline
(fig.~\ref{fig:lopo}). The reservoir is a
phenomenological rather than biophysical model, with fixed random recurrent
weights, and our ``stimulation'' is an idealised modification of the read-out or
its input rather than a literal physical protocol; the mapping between in-silico
amplitude and a deliverable dose is therefore approximate, and the anatomical
neighbourhoods rely on coarse parcel centroids. These factors bear on
quantitative details but not on the central, qualitative contrast between the
distributed and focal interventions, which is large and consistent across
classifiers and protocols.

Future work should test the distributed signature longitudinally and in larger,
multi-site cohorts, replace the phenomenological reservoir with more
biophysically grounded generative models, and connect the in-silico targets to
realisable stimulation montages so that distributed and multifocal strategies
can be evaluated explicitly. Formally testing whether the distributed FC
signature follows the predicted spread of pathology along network connections,
rather than merely co-localising with the most-affected hubs, would turn the
reconciliation proposed above into a falsifiable hypothesis.

\section*{Methods}

\paragraph{Data and preprocessing.}
Resting-state functional MRI sessions were downloaded from the Alzheimer's
Disease Neuroimaging Initiative (ADNI) database
(\href{https://adni.loni.usc.edu}{adni.loni.usc.edu}), comprising cognitively
unimpaired controls (CC) and patients with Alzheimer's disease (AD). Each session
comprised $140$ volumes (repetition time $\mathrm{TR}=\SI{3.0}{\second}$,
$\sim$\SI{7}{\minute}). Images were minimally preprocessed and parcellated into
$N_{\mathrm{p}}=121$ regions of interest, the $100$ cortical parcels of the
Schaefer 7-network atlas~\cite{schaefer2018} plus $21$ Harvard--Oxford
subcortical labels. All $121$ parcels (cortical and subcortical) enter the
reservoir fit, the classifiers and the stimulation analyses throughout; the
network-sorted group-mean FC matrices of Fig.~\ref{fig:data}D,E display the
$100$ cortical parcels only, for which the Yeo 7-network assignment is defined.
Region-wise BOLD time series were denoised by regression of $24$ head-motion
parameters and mean white-matter and cerebrospinal-fluid signals, and
band-pass filtered to $0.01$--$\SI{0.10}{\hertz}$. To balance the two groups
during reservoir fitting, classification and stimulation we sampled $40$ CC
sessions to match the $40$ AD subjects passing quality control; grouping by
subject (one session each) gives $36$ CC and $40$ AD subjects. The full pool
($145$ CC and $41$ AD subjects; $561$ and $151$ sessions) was used for the
cohort-size scaling and for the tangent-space SVM benchmark. The latter used the
globally signal-regressed sessions ($557$ CC $+$ $145$ AD sessions passing a
$>$$100$-volume filter) under \emph{nested} $5\times5$ subject-grouped
cross-validation: Ledoit--Wolf covariances projected to the tangent space at a
\emph{train-only} reference mean, then standardised, PCA-reduced, and classified
by a linear SVM or logistic regression, with the embedding dimension, penalty and
classifier chosen in the inner loop and evaluated on held-out subjects in the
outer loop.
Sessions were grouped by subject, and per-subject quantities used the first
available session unless otherwise stated; Table~\ref{tab:cohorts} summarises the
cohort used in every analysis.

\paragraph{Population principal-component projection.}
All region time series were pooled and mean-centred, and the leading
$N_{\mathrm{PC}}=50$ eigenvectors $\mathbf{V}_{50}$ of the
$N_{\mathrm{p}}\times N_{\mathrm{p}}$ covariance were retained. For each session
with signal $\mathbf{s}\in\mathbb{R}^{N_{\mathrm{p}}\times T}$ we defined the
PCA-reconstructed \emph{target}
$\hat{\mathbf{s}} = \mathbf{V}_{50}\mathbf{V}_{50}^{\!\top}\mathbf{s}$, which
removes idiosyncratic high-order variance while preserving the dominant
spatiotemporal structure. Empirical functional connectivity (FC) was the
Pearson correlation matrix of $\hat{\mathbf{s}}$. The $N_{\mathrm{PC}}=50$
projection defines only the reservoir's training \emph{target}; the read-out and
all FC quantities remain in the full $N_{\mathrm{p}}=121$ region space, so the
regional dimensionality is never reduced below $121$.

\paragraph{Reservoir model.}
We used a fixed random recurrent network of $N=2000$ rate units with state
$\mathbf{x}(t)$, internal weights $\mathbf{J}$ (Gaussian, rescaled to spectral
radius $0.95$), and fixed random input weights $\mathbf{J}_{\mathrm{in}}$
($\sigma_{\mathrm{in}}=0.01$; the subscript ``in'' distinguishes these untrained
input weights from the trained read-out $\mathbf{W}$). The network is a
\emph{linear}, deterministic reservoir: the update below is linear in
$\mathbf{x}$ and no endogenous noise is injected into the dynamics
(regularisation noise enters only the read-out fit, Eq.~2). Units evolved in
discrete time ($\mathrm{d}t=0.005$) as
\begin{equation}
\mathbf{x}(t+\mathrm{d}t) = e^{-\mathrm{d}t/\tau}\,\mathbf{x}(t)
  + \bigl(1-e^{-\mathrm{d}t/\tau}\bigr)
    \bigl(\mathbf{J}\,\mathbf{x}(t) + \mathbf{J}_{\mathrm{in}}\,\mathbf{u}(t)\bigr),
\end{equation}
with a short membrane time constant $\tau$ (effectively instantaneous update),
and a linear read-out $\mathbf{y}(t)=\mathbf{W}^{\!\top}\mathbf{x}(t)$ in the
$N_{\mathrm{p}}=121$-dimensional region space. The same reservoir (identical
$\mathbf{J},\mathbf{J}_{\mathrm{in}}$) was reused for every session.

\paragraph{Read-out fitting.}
For each session the reservoir was teacher-forced by feeding the target as
input, $\mathbf{u}(t)=f\,\hat{\mathbf{s}}(t)$ with feedback gain $f=0.1$, and
the hidden states $\mathbf{X}\in\mathbb{R}^{T_{\mathrm{eff}}\times N}$ were
collected after discarding the first $10$ steps. The read-out
$\mathbf{W}\in\mathbb{R}^{N\times N_{\mathrm{p}}}$ was obtained by
ridge-like regularised least squares with additive Gaussian noise of scale
$\sigma$ in the regressor,
\begin{equation}
\mathbf{W} = \bigl(\mathbf{X}+\mathbf{E}\bigr)^{+}\,\mathbf{Y},
\qquad E_{ij}\sim\mathcal{N}(0,\sigma^{2}),
\end{equation}
where $\mathbf{Y}$ is the time-aligned target and $(\cdot)^{+}$ the
Moore--Penrose pseudoinverse. The noise scale $\sigma$ acts as the
ridge-regularisation parameter of the read-out fit (larger $\sigma$, stronger
regularisation); $\sigma=0.025$ unless swept
(Fig.~\ref{fig:classification}).

\paragraph{Closed-loop reconstruction.}
The model has two operating modes. In \emph{teacher-forced} (open-loop) mode the
real target drives the input, $\mathbf{u}(t)=f\,\hat{\mathbf{s}}(t)$; this mode is
used to fit $\mathbf{W}$ and to compute the classifier features. In
\emph{closed-loop} (free-run) mode the read-out feeds back its own output,
$\mathbf{u}(t)=f\,\mathbf{y}(t)$, generating an autonomous reconstruction. To
validate the generative model (Fig.~\ref{fig:model}) the reservoir was reset and
run closed-loop: the real target drove the network for the first $5$ steps, after
which it free-ran for the remainder of the session. The
simulated reconstruction $\mathbf{Y}_{\mathrm{sim}}$ and its (lagged) FC were
compared to the empirical signals. Subject identifiability was quantified by
the matrix of correlations between simulated subject $i$ and empirical subject
$j$, comparing the same-subject (diagonal) and cross-subject (off-diagonal)
distributions. Both intervention families are evaluated consistently in both
modes: a read-out interpolation ($\mathbf{W}_{\mathrm{int}}$) and an input
oscillation ($A\sin$) each alter the same reservoir, and each is scored by the
classifiers on the teacher-forced FC-lag features \emph{and} assessed for its
departure from baseline on the closed-loop free-run FC
(Fig.~\ref{fig:singlecompare}, figs.~S5,~S8); the two are therefore compared on a
common footing, not one in open- and the other in closed-loop.

\paragraph{Read-out ``G-space''.}
Each read-out was projected onto the row space of its own reservoir states:
with $\mathbf{V}_{k}$ the top $K_{\mathrm{PC}}=200$ right singular vectors of
$\mathbf{X}$, the projected read-out is
$\widetilde{\mathbf{W}}=\mathbf{W}^{\!\top}\mathbf{V}_{k}\mathbf{V}_{k}^{\!\top}$,
vectorised to $\mathbf{w}_i$. Stacking patients, centring, and taking the SVD
of the centred matrix yielded patient coordinates (``G-scores'') in a
low-dimensional read-out-archetype space; the first $K$ components were used for
classification. The projection rank $K_{\mathrm{PC}}=200$ and the
low-dimensional sufficiency of the archetype basis are justified by a
training/test reconstruction-error analysis over $(K_{\mathrm{PC}},M)$
(fig.~S2).

\paragraph{Lagged-FC features.}
From the reconstructed signals $\mathbf{Y}=\mathbf{W}^{\!\top}\mathbf{X}$ we
computed lagged correlation matrices for lags $0$--$2$ (zero-lag upper triangle
plus full lag-$1$ and lag-$2$ matrices), concatenated into a feature vector per
patient. A Gram-matrix SVD across patients (the transductive embedding) gave the
``FC-lag'' coordinates, of which the first $K$ were used.

\paragraph{Classification.}
CC vs AD was classified with two algorithms on each embedding: a balanced
Fisher linear discriminant (LDA; majority class sub-sampled to the minority
within each training set) and a random forest (RF; $300$ trees, $\sqrt{\cdot}$
features, minimum leaf size $3$, balanced class weights). Performance was
estimated by repeated stratified $5$-fold cross-validation ($80/20$ train/test,
$10$ repeats), reporting balanced accuracy and the area under the ROC curve
(AUROC) as the mean $\pm$ standard deviation over folds. We swept the embedding
dimension $K\in[2,50]$ and the read-out noise $\sigma\in[0.003,0.5]$; reference
slices were chosen by the marginal-mean AUROC across classifiers and the
complementary axis. The SVD embeddings were built transductively (across all
patients) while the classifiers were trained and tested on disjoint folds.

\paragraph{In-silico stimulation.}
The model was used as a counterfactual testbed by shifting each AD read-out
toward the CC-mean read-out,
$\mathbf{W}_{\mathrm{int}} = (1-\alpha)\,\mathbf{W}_{\mathrm{AD}}
 + \alpha\,\bar{\mathbf{W}}_{\mathrm{CC}}$, applied either to all $121$ columns
(sites; full-W) or to focal subsets. Here $\alpha$ is the read-out interpolation
strength: $\alpha=0$ leaves the unmodified AD read-out, $\alpha=1$ replaces it
entirely by the CC-mean read-out $\bar{\mathbf{W}}_{\mathrm{CC}}$, and $\alpha>1$
is over-correction. We nonetheless explore $\alpha>1$ because the exact $\alpha=1$
replacement can be insufficient: the reconstruction is closed-loop, so the
reservoir state driven by the AD input keeps the FC partly AD-like even under the
CC read-out, and crossing the decision boundary can require pushing past
$\bar{\mathbf{W}}_{\mathrm{CC}}$ (cf.\ fig.~S5). ($\alpha$ is distinct from the
oscillation amplitude $A$ below.) Focal sites were ranked per patient by the
column norm of $\Delta\mathbf{W}=\bar{\mathbf{W}}_{\mathrm{CC}}-
\mathbf{W}_{\mathrm{AD}}$: the top-$5$, the top-$1$, and the top-$1$ site plus
its $10$ nearest neighbours in MNI space (parcel centroids from the atlas). For
each $\alpha$ the perturbed read-out was re-scored by both classifiers, and the
AD$\rightarrow$CC reclassification rate was the fraction crossing the CC/AD
decision boundary. We report a heuristic reference amplitude of $\alpha=4$ for
summary comparisons; it is a convenient common operating point, not a
biophysically calibrated limit, and the full dose-response is shown throughout.

\paragraph{Oscillatory stimulation.}
As an alternative focal protocol, a sinusoid $A\sin(2\pi f t)$ (amplitude $A$ in
input units, in the same units as $\mathbf{J}_{\mathrm{in}}\mathbf{u}$ and
distinct from the read-out interpolation strength $\alpha$ above) was injected as
additional input at the top-$k$ sites ($k=1,2,5$) during the teacher-forced
pass, with $\mathbf{W}$ unchanged; the oscillation propagates through the
recurrent dynamics and reshapes the reconstructed FC, which was re-scored by the
FC-lag classifier. Two candidate drive frequencies were compared: the leading
complex eigenmode of $\mathbf{J}$ (model-intrinsic, requiring the eigendecomposition
of $\mathbf{J}$) and the peak of the empirical power spectrum of the AD reservoir
states (data-driven, obtained by FFT without inspecting $\mathbf{J}$); the two
agree closely. Because $\mathbf{J}$ (and hence its eigenspectrum) is fixed and
shared across subjects, the resonance $f_{\mathrm{eig}}$ is a property of the model,
not a subject-specific fit. Frequencies were swept over $f\in[0,0.5]$ cycles/step
and amplitude up to \emph{supra-physiological} values, i.e.\ beyond the RMS of
the endogenous teacher-forcing drive (the target scaled by the feedback gain
$0.1$), so that the stimulated-to-baseline amplitude ratio
$\mathrm{RMS}_{\mathrm{pert}}/\mathrm{RMS}_{\mathrm{base}}>1$ (fig.~S10).
The state-level response of the linear
reservoir to a periodic drive is governed by the transfer function
$(\mathrm{i}\omega\mathbf{I}-\mathbf{J})^{-1}\mathbf{J}_{\mathrm{in}}\mathbf{b}$,
the same linear response as in~\cite{diantonio2026},
which is large only when the drive frequency is near an eigenmode \emph{and} the
spatial input pattern $\mathbf{b}$ overlaps that eigenmode; a single site thus
resonates cleanly at $f_{\mathrm{eig}}$, whereas the summed pattern of several
sites overlaps other modes and de-tunes the state-level resonance. Because the
FC-lag score is a nonlinear functional of the read-out of the states rather than
the excitation amplitude itself, this state resonance does not translate into a
correspondingly sharp multi-site frequency optimum: we verified that selecting
the five sites of maximal modal excitation
$|\mathbf{w}^{\mathsf H}\mathbf{J}_{\mathrm{in}}|$ (with $\mathbf{w}$ the leading
left eigenvector of $\mathbf{J}$) does \emph{not} recover an $f_{\mathrm{eig}}$
optimum. The clean spectral prescription is therefore specific to single-site
drive.

\paragraph{Leakage-free target selection.}
To test the therapy-target reclassification for classifier leakage we repeated
the FC-lag analysis under leave-one-AD-patient-out cross-validation. For each
held-out AD patient, the FC-lag embedding (a kernel PCA of the centred per-patient
feature vectors), the balanced Fisher LDA, its decision threshold, and the
site-selection criterion were all recomputed from the remaining patients (all CC
plus the other AD patients); none of the held-out patient's data entered the
classifier. The held-out patient's baseline and per-site resonantly-stimulated
FC-lag features were then projected into this train-only space by out-of-sample
kernel projection, its \emph{personalised} site chosen as the argmax of the
train-only score reduction over the $121$ sites, and the \emph{population-average}
site as the argmax of the mean reduction across the training AD patients;
reclassification was scored against the train-only threshold. Because the
embedding projection is a Gram/kernel form, each fold is exact linear algebra on
the base-patient Gram matrix, with the reservoir simulated once per
(patient,\,site). We report the \emph{net cure} rate: of the patients on the AD
side of the train-only boundary at baseline, the fraction that stimulation moves
to the CC side (fig.~\ref{fig:lopo}). The two targets bound the achievable
effect from opposite sides: the \emph{personalised} figure still reflects a
per-patient argmax over the $121$ candidate sites and should be read as an upper
bound on achievable personalisation, whereas the \emph{population-average}
result, a single global site selected on the training patients alone, is the
conservative, fully out-of-sample estimate.

\paragraph{Physiological diagnostics.}
For every strategy and $\alpha$ we recorded the signal-amplitude ratio at
stimulated sites, $\mathrm{RMS}_{\mathrm{pert}}/\mathrm{RMS}_{\mathrm{base}}$,
and the mean absolute FC at those sites, to verify that reclassification (where
it occurred) was achieved within plausible amplitude bounds.

\paragraph{Statistics and software.}
Group comparisons used two-sided Mann--Whitney $U$ tests; distributions are
shown as violins with medians or as mean $\pm$ SD/IQR as stated. Analyses were
implemented in Python (NumPy, scikit-learn, nilearn, Matplotlib).

\paragraph{Ethics.}
This work is a secondary analysis of de-identified, publicly available human
imaging data obtained from ADNI. The ADNI study was approved by the
institutional review boards of all participating sites, and all participants
(or their authorised representatives) provided written informed consent at
enrolment; no new data were acquired for the present study.

\paragraph{Data and code availability.}
The imaging data analysed here were downloaded from the Alzheimer's Disease
Neuroimaging Initiative (ADNI) database
(\href{https://adni.loni.usc.edu}{adni.loni.usc.edu}). ADNI data are available
to qualified researchers upon registration and acceptance of the ADNI Data Use
Agreement; in accordance with that agreement, the raw and derived imaging data
are not redistributed here. The exact cohorts and sample sizes used in every
analysis are listed in Table~\ref{tab:cohorts}. All analysis and
figure-generation code is openly available at
\href{https://github.com/cristianocapone/AD-reservoir-FC-stimulation}%
{github.com/cristianocapone/AD-reservoir-FC-stimulation}.

\paragraph{Acknowledgments.}
Data used in preparation of this article were obtained from the Alzheimer's
Disease Neuroimaging Initiative (ADNI) database
(\href{https://adni.loni.usc.edu}{adni.loni.usc.edu}). As such, the
investigators within the ADNI contributed to the design and implementation of
ADNI and/or provided data but did not participate in the analysis or writing of
this report. A complete listing of ADNI investigators can be found at
\href{https://adni.loni.usc.edu/wp-content/uploads/how_to_apply/ADNI_Acknowledgement_List.pdf}%
{adni.loni.usc.edu/wp-content/uploads/how\_to\_apply/ADNI\_Acknowledgement\_List.pdf}.
Data collection and sharing for ADNI was funded by the ADNI (National Institutes
of Health grant U01~AG024904) and DOD ADNI (Department of Defense award number
W81XWH-12-2-0012).
This work has been partially supported by the National Plan for Complementary
Investments to the NRRP, project ``D34H---Digital Driven Diagnostics, prognostics
and therapeutics for sustainable Health care'' (project code: PNC0000001), Spoke
3, funded by the Italian Ministry of University and Research.

% ============================================================================
\clearpage
% , , , , , , , , , , , , , , , , , , , , , , , , , --
%  References (BibTeX).  All entries live in refs.bib.
%  Only cited works are printed (\nocite{*} removed for submission).
%  naturemag.bst = Nature house reference style (superscript citations via
%  natbib [super]); verify every entry (volume/issue/pages/DOI) against the
%  publisher record before submission.
% , , , , , , , , , , , , , , , , , , , , , , , , , --
\bibliographystyle{naturemag}
\bibliography{refs}

% ============================================================================
\clearpage
\beginsupplement
\section*{Supplementary Information}

\subsection*{Supplementary Note S1: Structural versus functional discrimination}
On the same subjects and atlas, parcel-wise grey-matter volume classified AD from
controls at AUROC $\approx0.84$ (linear discriminant analysis) and generalised
across scanner sites (leave-one-site-out AUROC $0.64$--$0.71$), exceeding both
functional read-outs (FC-lag and empirical tangent-space FC, AUROC
$\approx0.62$--$0.66$, at chance across sites). Combining functional with
structural features did not improve on structure alone. This is consistent with
atrophy being the stronger \emph{diagnostic} signal; we nonetheless base the
intervention analysis on the functional read-out, for the reasons given in the
Introduction, because stimulation acts on dynamics rather than on tissue loss.

\begin{table}[htbp]
\centering
\small
\caption{\textbf{Cohorts used in each analysis.} Minimally preprocessed
resting-state fMRI ($121$ parcels, $140$ volumes, $\mathrm{TR}=3$\,s).
``Subjects'' are unique participants; ``sessions'' are individual scans (a subject
may contribute several). Per-subject reservoir quantities use the first valid
session.}
\label{tab:cohorts}
\begin{tabular}{p{4.9cm} c c p{3.4cm}}
\hline
Analysis & Controls (CC) & AD & Cross-validation / unit \\
\hline
Available pool & 145 subj / 561 ses & 41 subj / 151 ses & ,  \\
Data overview, FC similarity (Fig.~1) & 561 sessions & 151 sessions & per session \\
Reservoir fit \& FC validation (Fig.~2); G-space \& FC-lag classifiers
(Fig.~3A--D); all in-silico stimulation (Figs~4--6, S2--S10)
& 36 subjects & 40 subjects & one session/subject; $40$ CC sessions sampled to
balance $40$ AD ($36$ unique CC subjects); repeated stratified CV \\
Cohort-size scaling (Fig.~S1) & up to 145 subj & up to 41 subj & balanced subsets,
leave-one-patient-out \\
Tangent-space SVM benchmark (Fig.~3E,F) & 557 sessions (143 subj) & 145 sessions
(40 subj) & GSR preprocessing; nested $5\times5$ subject-grouped CV ($702$
sessions, train-only tangent reference) \\
\hline
\end{tabular}
\end{table}

\begin{figure}[htbp]
  \centering
  \includegraphics[width=\linewidth]{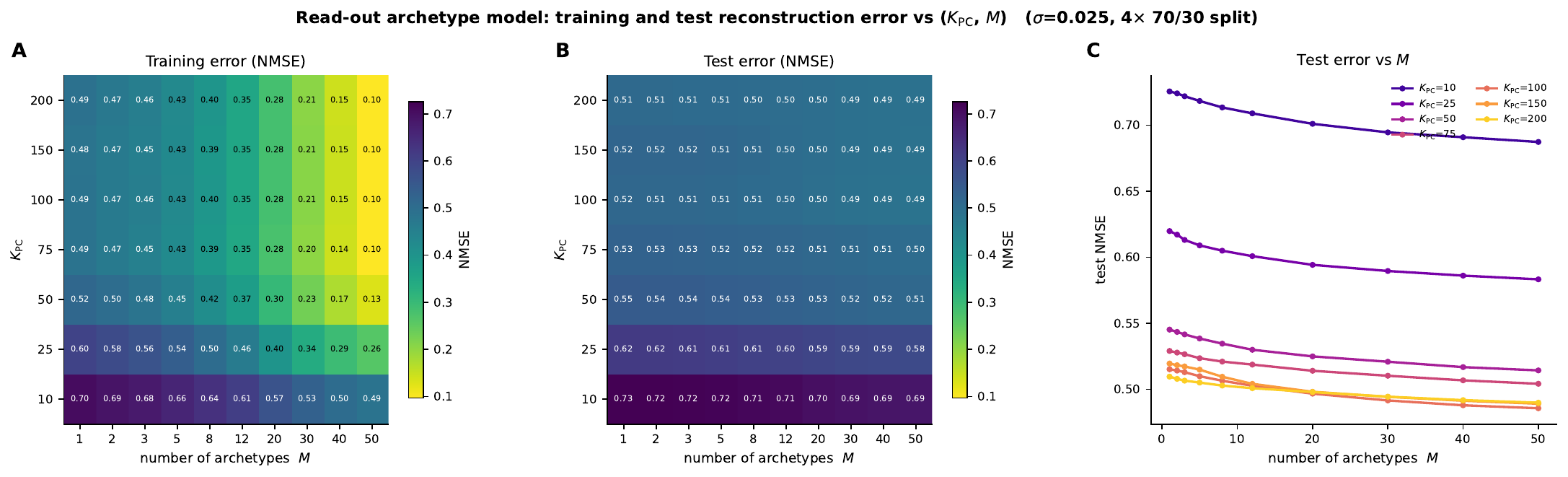}
  \caption{\textbf{Training and test reconstruction error of the read-out
  archetype model.}
  Each fitted read-out $\mathbf{W}_i$ is projected onto the top-$K_{\mathrm{PC}}$
  right singular vectors of its reservoir states, and the projected read-outs are
  then compressed across patients into $M$ SVD archetypes; the error is the
  signal-space normalised mean-squared error,
  $\mathrm{NMSE}=\langle(\mathbf{Y}-\mathbf{X}\mathbf{W}_g^{\!\top})^2\rangle/
  \langle\mathbf{Y}^2\rangle$, of the archetype reconstruction $\mathbf{W}_g$
  ($\sigma=0.025$; mean over $4$ random $70/30$ patient splits).
  (\textbf{A}) Training and (\textbf{B}) test NMSE over the
  $(K_{\mathrm{PC}},M)$ grid.
  (\textbf{C}) Test NMSE vs $M$ for each $K_{\mathrm{PC}}$.
  Training error falls steeply with both $K_{\mathrm{PC}}$ and $M$, whereas test
  error is governed chiefly by $K_{\mathrm{PC}}$ (plateauing for
  $K_{\mathrm{PC}}\gtrsim100$) and is nearly flat in $M$ beyond $M\!\approx\!10$:
  additional archetypes fit training-specific variance, the widening
  train--test gap, without improving generalisation. This justifies the
  projection rank $K_{\mathrm{PC}}=200$ used throughout and shows that a
  low-dimensional archetype basis captures the cross-patient-shared read-out
  structure.}
  \label{fig:archetype}
\end{figure}

\begin{figure}[htbp]
  \centering
  \includegraphics[width=\linewidth]{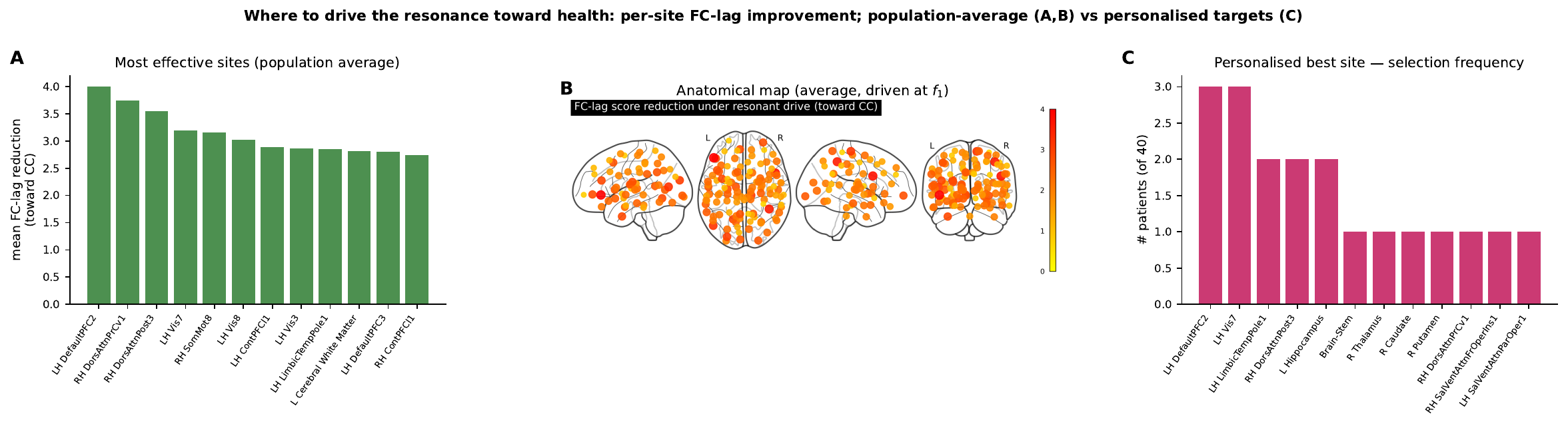}
  \caption{\textbf{Where to drive the resonance toward health.}
  Each of the $121$ sites is driven at the leading eigenmode frequency $f_1$
  (amplitude $A=4$) and ranked by the resulting FC-lag LDA-score reduction toward
  CC, i.e.\ the projection of the resonant-stimulation effect onto the
  discriminant.
  (\textbf{A}) Population-average ranking of the most effective sites; the leading
  target is the left Default-network prefrontal cortex, followed by
  dorsal-attention, visual and somatomotor association cortex.
  (\textbf{B}) The same on a glass brain (marker size/colour = mean reduction):
  the effective sites are predominantly cortical and distributed.
  (\textbf{C}) When the target is chosen \emph{per patient} (each subject's own
  model), the selected sites are heterogeneous, most are chosen by only a few
  patients and span cortical (Default, visual) and subcortical (hippocampus,
  brainstem, thalamus) structures, so no single global target is optimal for
  all, motivating personalised targeting.}
  \label{fig:ldares}
\end{figure}

\begin{figure}[htbp]
  \centering
  \includegraphics[width=\linewidth]{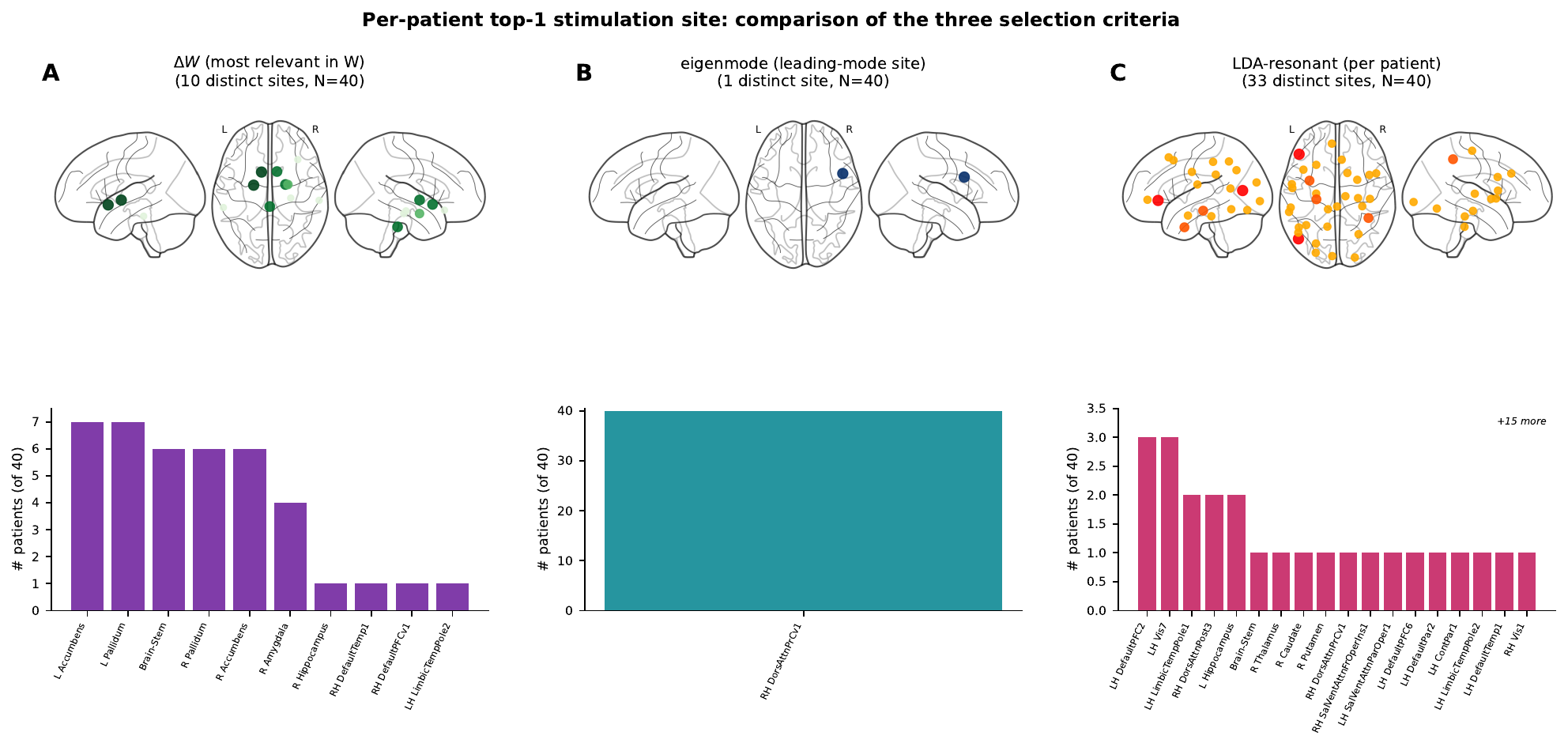}
  \caption{\textbf{Per-patient top-1 stimulation site: comparison of the three
  selection criteria} (same seeded reservoir; glass brain on top, selection-%
  frequency histogram below; $N=40$ AD patients).
  (\textbf{A}) \emph{$\Delta W$} (most relevant in the read-out; the single-site
  $\Delta W$ stimulation): the site of largest $\|W_{\mathrm{CC}}-W_p\|$. These
  targets cluster on a few \emph{subcortical} structures shared across patients
  ($10$ distinct sites; mostly accumbens, pallidum, brain-stem, amygdala), i.e.\
  where the disease most distorts the read-out.
  (\textbf{B}) \emph{Eigenmode}: the site of maximal coupling to the leading
  reservoir eigenmode. Being a property of the (shared) dynamics, it is a
  \emph{single global site} for all patients (right dorsal-attention cortex).
  (\textbf{C}) \emph{LDA-resonant} (per patient): the site whose resonant drive
  moves that subject's FC-lag discriminant most toward CC. These therapeutic
  targets are highly heterogeneous, \textbf{$33$ distinct sites for the $40$
  patients}, each selected by at most three, spanning Default-network, visual and
  attention cortex together with medial-temporal and subcortical structures.
  The three criteria answer different questions: where the disease lives in the
  read-out ($\Delta W$), the fixed resonance node of the dynamics (eigenmode), and
  the personalised lever that drives recovery (LDA-resonant); only the last is
  genuinely individualised, so no single global site is adequate.}
  \label{fig:persdist}
\end{figure}

\begin{figure}[htbp]
  \centering
  \includegraphics[width=\linewidth]{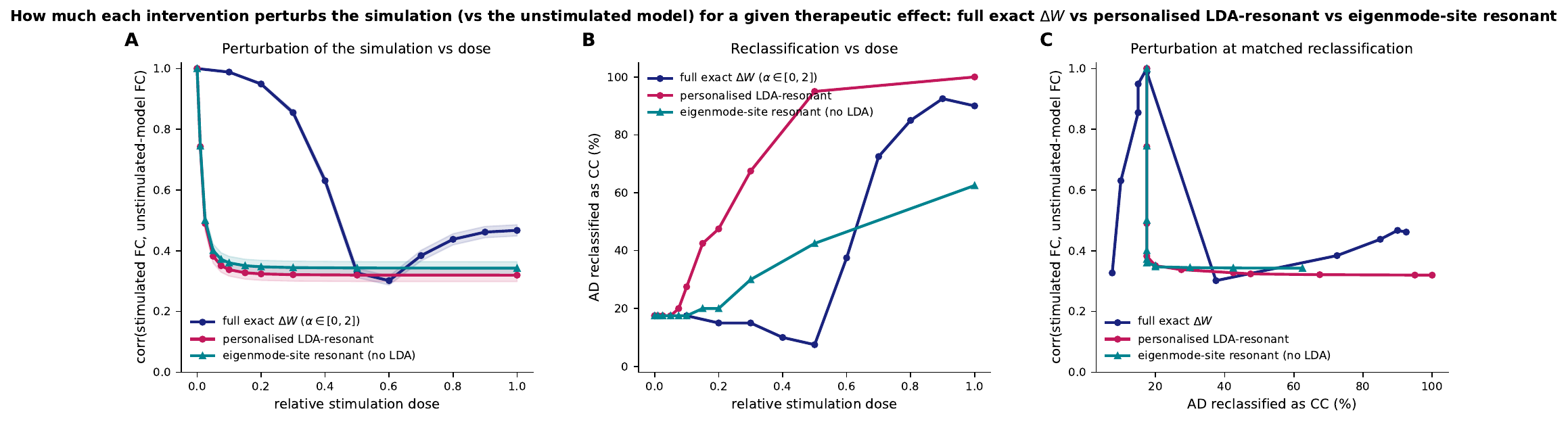}
  \caption{\textbf{How much each intervention perturbs the simulation, for a given
  therapeutic effect: full exact $\Delta W$, personalised LDA-resonant, and
  eigenmode-site resonant stimulation} ($N=40$ AD). The perturbation is the Pearson
  correlation (FC off-diagonals) between the \emph{closed-loop free-run} simulated
  FC under stimulation (generative protocol of Fig.~\ref{fig:model}: drive $5$
  steps, then free-run) and the \emph{unstimulated} model's own free-run FC, so it
  equals $1$ at zero dose by construction and falls as the intervention reshapes the
  dynamics. (The unstimulated free-run reproduces the empirical FC at
  $r\!\approx\!0.61$, as in Fig.~\ref{fig:model}.) Three interventions are compared:
  full-site $\Delta W$ interpolation $W_{\mathrm{int}}=(1-\alpha)W_p+\alpha
  W_{\mathrm{CC}}$ ($\alpha\in[0,2]$; $\alpha=1$ full replacement by the CC-mean
  read-out, $\alpha>1$ over-correction), the \emph{personalised} LDA-resonant drive
  (each patient's own discriminant-aligned site), and the \emph{eigenmode-site}
  resonant drive (single global site of maximal coupling to the leading eigenmode,
  selected by modal coupling \emph{without} the LDA); both oscillatory drives use
  $A\in[0,10]$ at $f_1$, sampled finely at low $A$. (\textbf{A}) Perturbation versus
  normalised dose. Both single-site resonant drives fall \emph{steeply but
  continuously} ($1\!\to\!0.74\!\to\!0.49\!\to\!0.38$ over $A=0.1$--$0.75$) and
  saturate near $r\!\approx\!0.33$: because the injected $A\sin$ is added to the
  un-scaled input whereas the data drive is $\mathit{ff}\!\cdot\!\mathrm{tgt}$
  ($\mathit{ff}=0.1$), even $A=1$ is an order of magnitude above the normal drive,
  and at $f_1$ the closed loop amplifies it until the resonant mode's
  (amplitude-normalised) spatial pattern dominates the FC. This perturbation is
  \emph{network-wide}, not local to the driven node: recomputing the correlation
  after deleting the driven site's own row and column leaves it unchanged
  ($r=0.32$ vs $0.32$ at $A=10$), i.e.\ every parcel, not just the stimulated
  one, becomes phase-locked to the driven mode. (\textbf{B})
  Reclassification versus dose: the full $\Delta W$ correction reverts the FC-lag
  classifier only with over-correction ($\alpha\!\gtrsim\!1.4$; $\le$$8\%$ at the
  exact $\alpha=1$, since the reservoir state stays AD-driven, consistent with
  Fig.~\ref{fig:stim}), reaching $\sim$$90\%$ by $\alpha=2$; the personalised
  LDA-resonant drive reaches $100\%$, whereas the eigenmode-site drive plateaus at
  $\sim$$62\%$. (\textbf{C}) Perturbation at matched reclassification. The two
  single-site resonant drives perturb the simulation \emph{near-identically}
  (correlation to the unstimulated model $\approx0.32$--$0.34$ throughout), yet only
  the discriminant-aligned (LDA) target converts that same dynamical perturbation
  into full reclassification, selecting the site by its effect on the discriminant,
  rather than by modal coupling alone, extracts more therapeutic effect per unit of
  perturbation. The global $\Delta W$ overwrite perturbs less at low effect but must
  over-correct and rewrite all $121$ read-out weights to compete.}
  \label{fig:fidelity}
\end{figure}

\begin{figure}[htbp]
  \centering
  \includegraphics[width=\linewidth]{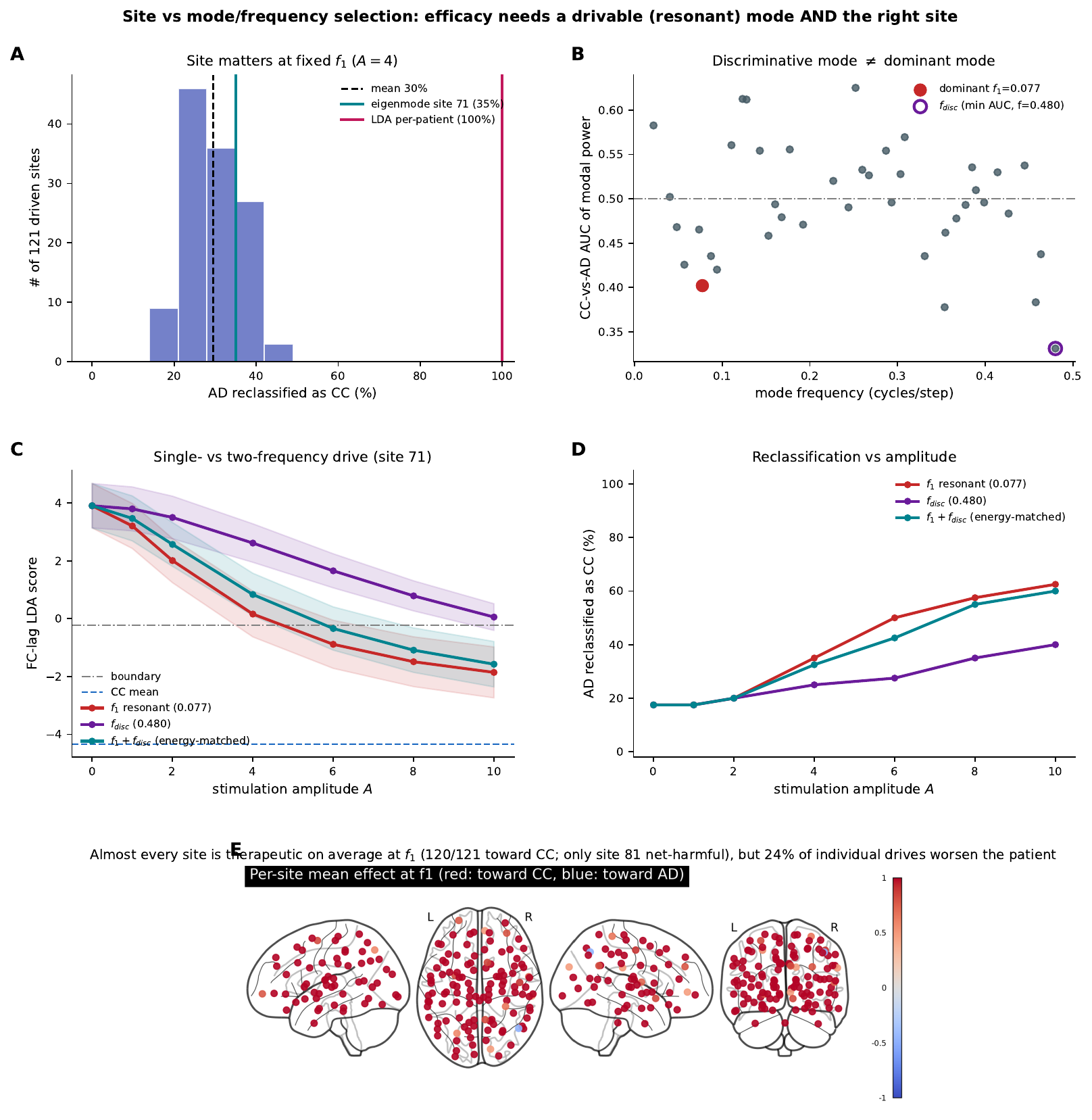}
  \caption{\textbf{Why site and resonance both matter: a drivable mode is not the
  same as a disease-discriminative one} ($N=40$ AD). Because a resonant drive
  reshapes the FC network-wide (fig.~S5), one might hope that the frequency/mode
  alone determines the effect. It does not. (\textbf{A}) Driving each of the $121$
  sites at the dominant-mode frequency $f_1$ ($A=4$): the per-site reclassification
  rate spans the whole low range ($15$--$48\%$, mean $30\%$; no site reaches
  $\ge$$90\%$), so the \emph{site} still decides the outcome, only per-patient
  (LDA) selection reaches $100\%$. (\textbf{B}) Ranking eigenmodes by how well their
  per-patient modal power separates CC from AD (AUC): the most discriminative modes
  lie at higher frequencies ($f\!\approx\!0.25$--$0.48$), while the dominant
  (least-damped, most \emph{drivable}) mode $f_1=0.077$ is only weakly
  discriminative (AUC $0.40$, rank $7/40$). (\textbf{C},\textbf{D}) At a fixed site
  (eigenmode site $71$), driving the most discriminative \emph{deficit} mode
  $f_{\mathrm{disc}}=0.48$ (largest AD power deficit, the only correctable
  direction since drive can only add power) is markedly \emph{less} effective than
  the resonant $f_1$ ($40\%$ vs $62\%$ reclassified at $A=10$), and an
  energy-matched two-frequency drive $f_1+f_{\mathrm{disc}}$ does \emph{not} beat
  $f_1$ alone ($60\%$ vs $62\%$): diverting energy to the weakly-driven
  discriminative mode loses more amplification than it gains in direction. Efficacy
  thus requires a \emph{drivable} (resonant) mode \emph{and} the right site, which
  is precisely what the LDA-resonant criterion optimises, and why it outperforms
  modal-coupling, disease-deviation, or discriminative-frequency heuristics.
  (\textbf{E}) Signed per-site mean effect at $f_1$ on a glass brain (red: toward
  CC; blue: toward AD). Almost every site is therapeutic on average
  ($120/121$ toward CC; only the right control/parietal site is net-harmful, and
  only marginally). This is \emph{not} because the drive finds the CC direction by
  chance: a resonant drive injects the \emph{same} stereotyped dominant-mode pattern
  into every patient (site-independent; fig.~S5), so there is effectively one driven
  direction, and it is CC-aligned because (i) AD sits at the far extreme of the
  discriminant (baseline score $+3.9$ vs boundary $-0.22$, CC mean $-4.3$) and the
  low-structure oscillatory FC lacks the AD-specific lag pattern, and (ii) the
  dominant mode is mildly under-expressed in AD (panel B, AUC $0.40$), so adding its
  power is restorative. The effect is a directed \emph{shift}, not a collapse, it
  moves the score mean $\sim$$5.8$ toward CC while preserving the per-patient spread
  ($\pm4.9\!\to\!\pm5.6$), so a patient-specific minority still worsens ($24\%$ of
  all (site,patient) drives push toward AD, flipping $5/40$ borderline patients back
  across the boundary), a safety argument for personalised targeting. The
  ``therapeutic'' action is thus largely a regularisation that erases AD-specific
  structure rather than a precise restoration of healthy connectivity.}
  \label{fig:modefreq}
\end{figure}

\begin{figure}[htbp]
  \centering
  \includegraphics[width=\linewidth]{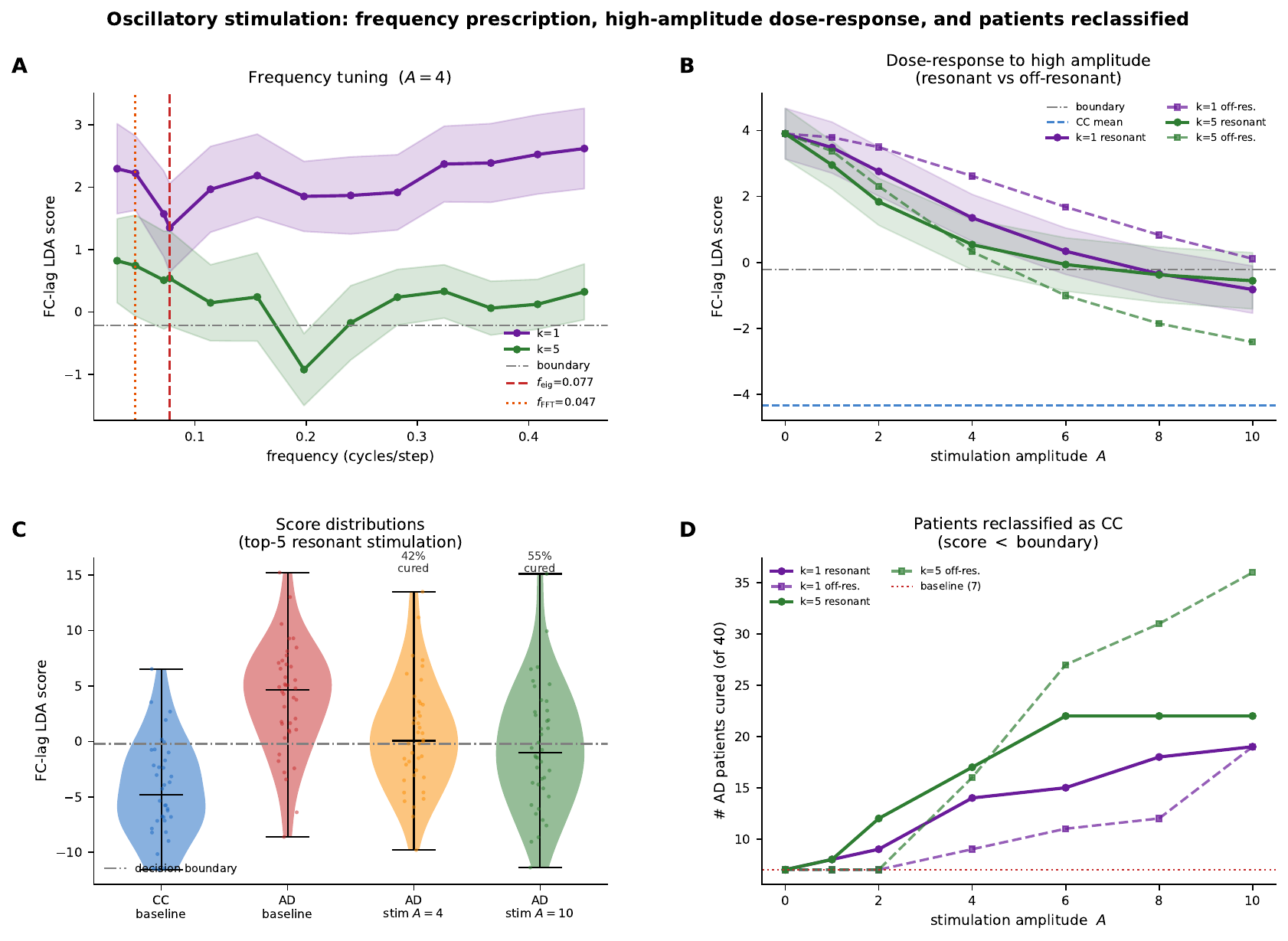}
  \caption{\textbf{Frequency prescription and high-amplitude dose-response of
  focal oscillatory stimulation} ($N=40$ AD). A sinusoidal drive
  $A\sin(2\pi f t)$ is injected at the top-$k$ most-affected ($\Delta W$) sites and
  the FC-lag score re-evaluated (lower\,$=$\,more control-like).
  (\textbf{A}) Frequency tuning at moderate amplitude ($A=4$): mean $\pm$ SEM for
  the single top site ($k=1$, purple) and the top-five sites ($k=5$, green). For
  $k=1$ the score is minimal precisely at the reservoir's dominant eigenmode
  frequency $f_{\mathrm{eig}}=0.077$ (red dashed), a sharp resonance;
  $f_{\mathrm{FFT}}$ (orange dotted) marks the spectral peak of the AD reservoir
  states. The $k=5$ optimum broadens and shifts, as the FC-lag effect is a
  nonlinear functional of the dynamics, not the modal excitation amplitude.
  (\textbf{B}) Dose-response to high amplitude ($A=10$), resonant
  ($f_{\mathrm{eig}}$, solid) vs off-resonance (highest frequency, dashed) for
  $k=1$ and $k=5$; scores move toward the CC mean (blue dashed) as amplitude grows.
  (\textbf{C}) FC-lag score distributions: CC baseline, AD baseline, and AD after
  top-five resonant stimulation at $A=4$ and $A=10$ (decision boundary dot-dashed;
  fraction crossing it annotated).
  (\textbf{D}) Patients reclassified as CC vs amplitude, for
  top-1/top-5 $\times$ resonant/off-resonance (baseline dotted): resonance is most
  efficient at low-to-moderate amplitude, whereas at supra-physiological amplitude
  even an off-resonance multi-site drive reclassifies most patients
  ($36/40$ at $A=10$), an energy-driven (non-specific) saturation effect.}
  \label{fig:oscfreq}
\end{figure}

\begin{figure}[htbp]
  \centering
  \includegraphics[width=\linewidth]{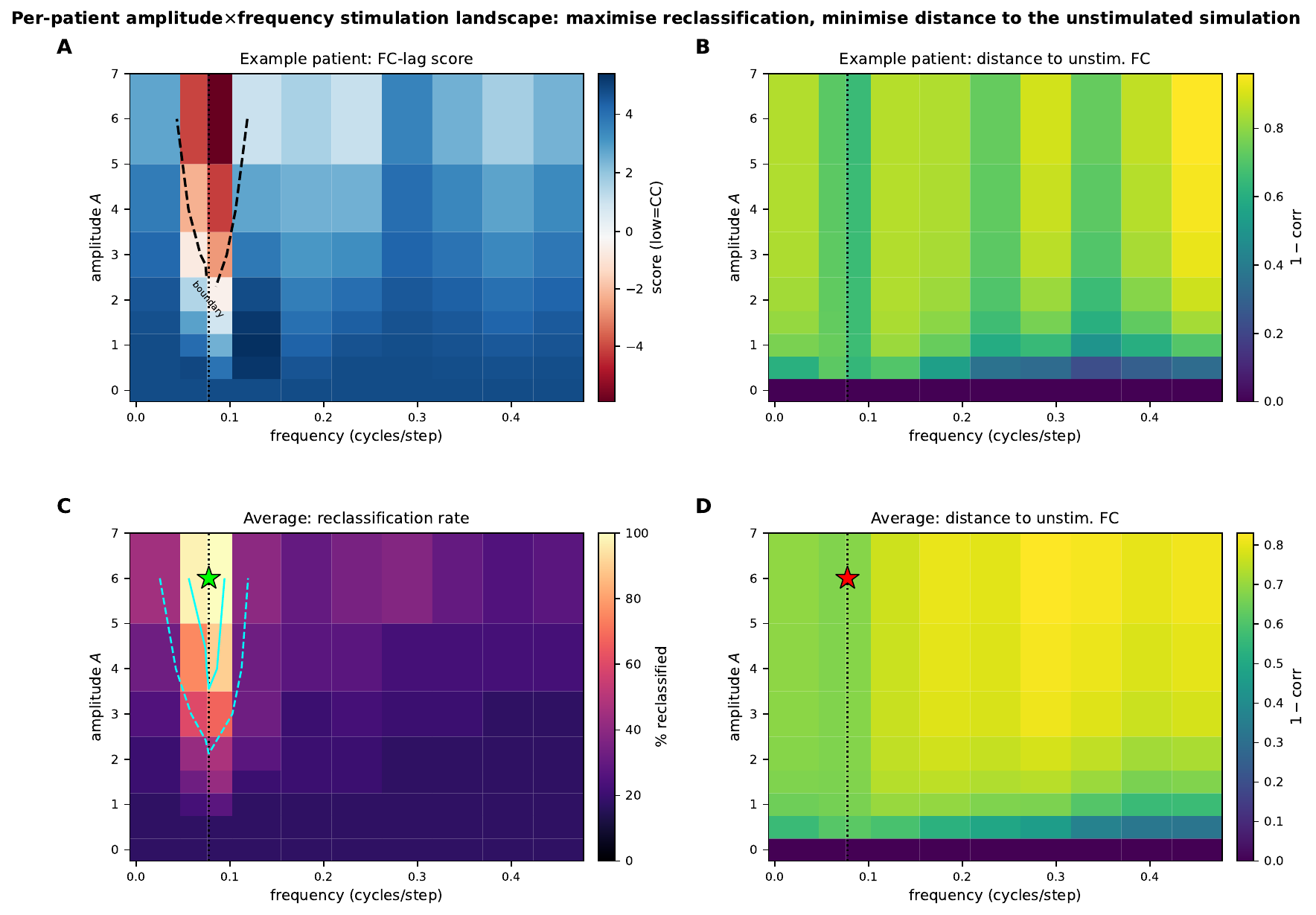}
  \caption{\textbf{Per-patient amplitude$\times$frequency stimulation landscape:
  trading off reclassification against departure from the unstimulated simulation}
  ($N=40$ AD). Each patient is driven at its personalised LDA-resonant site over an
  $8\times10$ grid of amplitude $A$ and frequency $f$; we record the FC-lag score
  (reclassified if below the boundary; classifier space) and the distance to the
  \emph{unstimulated} simulation, $1-\mathrm{corr}$ between the stimulated and
  unstimulated closed-loop free-run FC.
  (\textbf{A},\textbf{B}) One representative patient (median baseline score): the
  FC-lag score (A; boundary contour in black) and the distance (B). The score
  drops below the boundary only in a narrow band of frequencies around the
  reservoir resonance $f_1$ (dotted line).
  (\textbf{C}) Population reclassification rate: a narrow ``tongue'' centred on
  $f_1$ (cyan contours at $50\%$ and $80\%$), at resonance patients reclassify at
  the \emph{lowest} amplitude, whereas off-resonance requires much larger drive.
  (\textbf{D}) Population mean distance, which grows with amplitude and is itself
  elevated near $f_1$. The two objectives therefore conflict: the operating point
  that maximises reclassification minus distance ($\star$; $A\!=\!6$, $f\!=\!f_1$,
  $100\%$ reclassified) sits at high distance ($\approx0.68$), and reverting most
  patients unavoidably requires a large perturbation (distance $\gtrsim0.66$, cf.\
  fig.~S5--S6). Resonance is nonetheless the most \emph{efficient} axis, the
  minimum-distance route to any target reclassification runs along $f_1$, so the
  model's prescription is to drive the personalised site at the intrinsic
  resonance, accepting the minimal perturbation that crosses the boundary.}
  \label{fig:ampfreq}
\end{figure}

\begin{figure}[htbp]
  \centering
  \includegraphics[width=\linewidth]{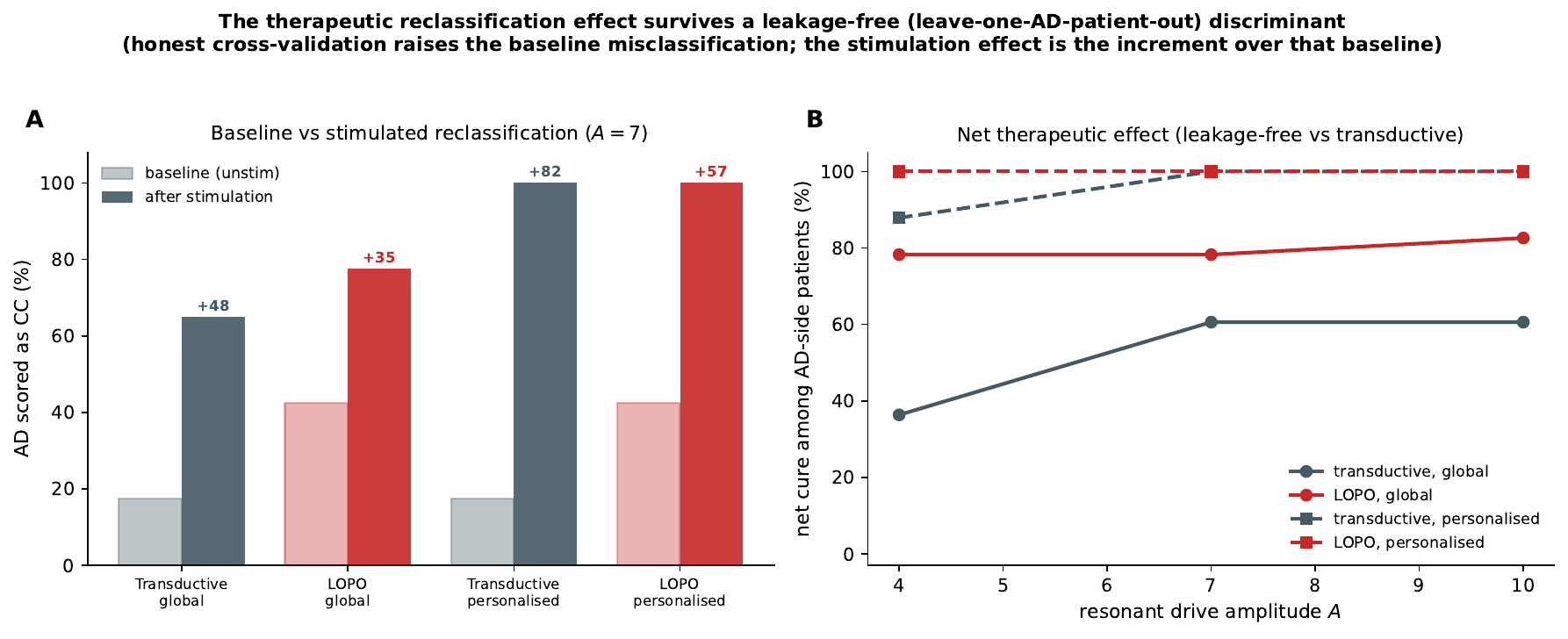}
  \caption{\textbf{The therapeutic reclassification survives a leakage-free
  (leave-one-AD-patient-out) discriminant} ($N=40$ AD). The FC-lag embedding,
  the LDA discriminant, its decision threshold and the site-selection criterion
  are all rebuilt from the training patients only and applied to each held-out
  AD patient (Methods). Two targets are compared: the
  \emph{population-average} site (one global site, selected on the training AD
  patients; conservative, no per-test-patient choice) and the \emph{personalised}
  per-patient LDA-resonant site (an upper bound, a per-patient argmax over the
  $121$ sites).
  (\textbf{A}) Fraction of AD patients scored as CC at baseline (unstimulated)
  versus after resonant stimulation at the reference amplitude $A=7$, for each
  target under the transductive scheme (fit on all patients) and the leakage-free
  LOPO scheme (train-only); increments annotated. Honest cross-validation raises
  the baseline misclassification of AD as CC from $17.5\%$ (transductive) to
  $42.5\%$ (LOPO), so that only $23$ of $40$ AD patients lie on the AD side of the
  boundary unstimulated.
  (\textbf{B}) Net cure, the fraction of patients on the AD side at baseline that
  stimulation moves to the CC side, versus resonant drive amplitude, transductive
  vs LOPO, for the population-average (solid) and personalised (dashed) targets.
  Leakage-free, the single population-average target reaches $\sim$$78\%$ net cure
  and the personalised target $\sim$$100\%$: the reclassification is a genuine,
  model-prescribed intervention effect, not an artefact of classifier leakage.}
  \label{fig:lopo}
\end{figure}

\begin{figure}[htbp]
  \centering
  \includegraphics[width=\linewidth]{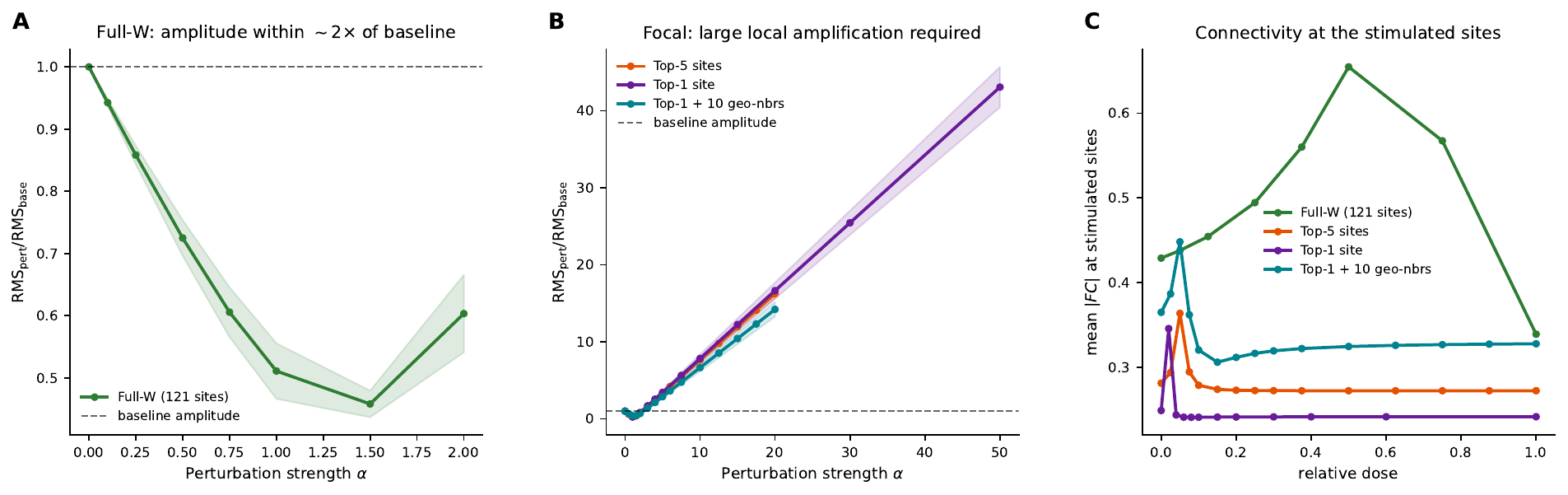}
  \caption{\textbf{Physiological diagnostics of the read-out interpolation:
  distributed correction stays near the endogenous amplitude, focal correction
  does not} ($N=40$ AD). At the stimulated sites we record the signal-amplitude
  ratio $\mathrm{RMS}_{\mathrm{pert}}/\mathrm{RMS}_{\mathrm{base}}$ between the
  perturbed reconstruction $Y=\mathbf{X}\mathbf{W}_{\mathrm{int}}$ and the
  unperturbed one, and the mean $|FC|$ at those sites (Methods,
  ``Physiological diagnostics''); mean $\pm1\sigma$ over AD patients.
  (\textbf{A}) Full-W correction: the amplitude ratio remains within a factor of
  about two of baseline over the whole range $\alpha\in[0,2]$ (dashed line,
  baseline), i.e.\ the distributed correction is achieved without driving the
  signal outside its endogenous range.
  (\textbf{B}) Focal corrections: the ratio instead grows approximately linearly
  with $\alpha$, reaching $\sim$$16\times$ (top-5) and $\sim$$44\times$ (top-1)
  baseline at the amplitudes needed for even partial effects, i.e.\ deep into
  supra-physiological territory.
  (\textbf{C}) Mean $|FC|$ at the stimulated sites versus relative dose.
  Together these confirm that the failure of the focal strategies
  (Fig.~\ref{fig:stim}C,F) is not a matter of insufficient dose: they are
  already operating far beyond plausible amplitude bounds.}
  \label{fig:physio}
\end{figure}

\end{document}